\documentclass[11pt,a4paper]{article}
\usepackage[utf8x]{inputenc}
\usepackage{graphicx}
\usepackage{grffile} 
\usepackage{amsmath,amssymb,amsbsy,color}
\usepackage{multirow}
\usepackage{hyperref}
\usepackage[numbers,compress]{natbib}
\graphicspath{{figures/}}

\begin{document}

\begin{flushright}
\small{
MITP/16-036\\
HIM-2016-01\\
}
\end{flushright}
\vspace{5mm}

\begin{center}
{\Large\bf Nucleon matrix elements from lattice QCD
with all-mode-averaging and a domain-decomposed solver:
an exploratory study}
\end{center}
\vspace{5mm}

\begin{center}
Georg von Hippel$^{(a)}$, Thomas D.~Rae$^{(b)}$, Eigo~Shintani$^{(a,c)}$, Hartmut Wittig$^{(a,d)}$\\[4mm]
{\small\it
$^a$PRISMA Cluster of Excellence and Institut f{\"u}r Kernphysik,\\
Johannes Gutenberg-Universit{\"a}t Mainz, D-55099 Mainz, Germany}\\
{\small\it
$^b$Bergische Universit{\"a}t Wuppertal, Gau\ss stra\ss e 20, D-42119~Wuppertal, Germany}\\
{\small\it
$^c$RIKEN Advanced Institute for Computational Science,
K\=obe, Hy\=ogo 650-0047, Japan}\\
{\small\it
$^d$Helmholtz Institute Mainz, University of Mainz, 55099 Mainz, Germany}
\\[10mm]
\end{center}

\begin{abstract}
We study the performance of all-mode-averaging (AMA) when used in
conjunction with a locally deflated SAP-preconditioned solver, determining
how to optimize the local block sizes and number of deflation fields in
order to minimize the computational cost for a given level of overall
statistical accuracy. We find that AMA enables a reduction of the statistical
error on nucleon charges by a factor of around two at the same cost when
compared to the standard method.
As a demonstration, we compute the axial, scalar and tensor charges of the
nucleon in $N_f=2$ lattice QCD with non-perturbatively O($a$)-improved Wilson
quarks, using O(10,000) measurements to pursue the signal out to source-sink
separations of $t_s\sim 1.5$~fm. Our results suggest that the axial charge
is suffering from a significant amount (5-10\%) of excited-state contamination
at source-sink separations of up to $t_s\sim 1.2$~fm, whereas the
excited-state contamination in the scalar and tensor charges seems to be small.
\end{abstract}

\section{Introduction}

Recent developments in the field of numerical algorithms and computer
hardware have made it possible to perform simulations of lattice QCD
with dynamical light quarks at physical pion masses, enabling
reliable first-principle determinations of hadronic and nuclear properties.

Indeed, with current technology the lattice computation of the light hadron
spectrum, including not only the ground states, but also resonances, has
become a routine task, which can be accomplished with great precision
\cite{Durr:2008zz,Mahbub:2009aa,Bulava:2010yg,Edwards:2012fx,
Alexandrou:2013fsu,Borsanyi:2014jba}.
On the other hand, attempts at achieving a similar precision for the
prediction of nucleon structure observables from lattice QCD
\cite{Martinelli:1988rr,Draper:1989pi,Gusken:1989ad,Leinweber:1990dv,
Yamazaki:2009zq,Syritsyn:2009mx,Capitani:2012gj,Horsley:2013ayv,
Bali:2014nma,Abdel-Rehim:2015jna}
are confronted with a dilemma arising from the simultaneous problems
of excited-state contamination at short, and deteriorating
signal-to-noise ratio at large Euclidean times:
while excited states are exponentially suppressed at large time separations,
the signal-to-noise ratio likewise decays exponentially with time,
and increases only with the square root of statistically
independent measurements. As a result, controlling both statistical and
systematic errors for nucleonic observables becomes difficult,
in particular for structure observables, where both the time separation 
between the nucleon source and sink, and those between the source or sink
and the operator insertion of interest, need to be made large to suppress
excited-state contaminations.

It is therefore perhaps not surprising that the results for nucleon
structure observables which have been obtained by different lattice
collaborations currently show large discrepancies between the different
groups (for details, cf. e.g.
\cite{Syritsyn:2014saa,Bhattacharya:2013ehc,Constantinou:2014tga,Green:2014vxa}
and references therein).
The problem is particularly acute for the axial charge of the nucleon,
which is both of fundamental importance for testing the limits of the Standard
Model and well known experimentally, but for which  lattice results differ
amongst themselves by several standard deviations and show a discrepancy
of about 10\% from the experimental value when extrapolated to the
physical point.

To address this problem, lattice studies of nucleon structure with
high statistical accuracy at large Euclidean time separations are
needed. Keeping the computational cost manageable requires efficient
techniques of variance reduction.
The present study aims to reduce the statistical noise
by using the recently proposed technique of all-mode-averaging (AMA)
\cite{Blum:2012uh,Blum:2012my,Shintani:2014vja}.
AMA is able to achieve a significant reduction in statistical error
at moderate cost by combining multiple cheap low-precision calculations
of the quark propagator with an appropriate bias correction.
This makes AMA particularly attractive for approaching the physical
light quark mass, since the cost of computing quark propagators scales
inversely proportional to the quark mass.

In this paper, we study the efficiency of AMA when combined with the
highly efficient locally deflated SAP-preconditioned GCR solver.
In an extension of our previous studies
\cite{Capitani:2012gj,Capitani:2011fg,Jager:2013kha},
we apply AMA to the calculation of the axial charge of the nucleon
from three-point functions with large source-sink separations of 
around $1.5$~fm and above
on large lattices satisfying $m_\pi L>4$ with $N_{\rm f}=2$
flavours of dynamical quarks.
In addition, we also determine the scalar and tensor charges of the nucleon
on the same configurations.

We find that for the axial charge as extracted from ratios of correlation
functions, large source-sink separations are
required to reliably suppress excited-state contaminations, which become
visible only at high enough statistics, and that values close to the
experimental one are obtained from the largest source-sink separations
studied. The summation method
\cite{Gusken:1989ad,Capitani:2012gj,Jager:2013kha}
is able to extract the asymptotic behaviour
already from moderate source-sink separations, but still
profits greatly from having precise measurements at large separations.

This paper is organized as follows:
In section \ref{sec:method}, we explain the numerical methods,
including how to properly define AMA when using the
Schwartz alternating procedure (SAP) and local deflation
with the GCR solver. We also define the ratios of three- and
two-point functions that we use to extract the axial charge,
scalar and tensor charge.
In section \ref{sec:performance}, we study the performance of AMA
and consider how to tune the solver parameters.
In Section \ref{sec:2pt} and \ref{sec:3pt}, we present a first
analysis of the nucleon two-point function and three-point 
functions, respectively, using AMA with parameters tuned as 
presented in Section \ref{sec:performance}.
In the last section, we summarize our results and discuss directions
for further improvement and future study.

\begin{table}[t]
\begin{center}
\caption{Lattice parameters and gauge ensembles used in this analysis. 
The number $N_{\rm meas}$ of measurements is given by $N_G$ (see Table~\ref{tab:param}) multiplied by
the number $N_{\rm conf}$ of configurations.}
\label{tab:latparam}
\begin{tabular}{cccccccc}
\hline\hline
Label & Lattice($T\times L^3$) & $a$ [fm] & $m_\pi$ [MeV] & $t_s$ [fm] & $N_{\rm conf}$ & $N_{\rm meas}$\\
\hline
A5 & $64\times 32^3$ & 0.079 & 316 & 0.79 & 68 & 4,352 \\
     & (2.5 fm)$^3$ & & ($m_\pi L=4.0$) & 0.95 & 74 & 4,672 \\
     & & & & 1.11 & 72 & 4,608\\
     & & & & 1.26 & 71 & 4,544\\
     & & & & 1.42 & 695 & 44,480\\
\hline
B6 & $96\times 48^3$ & 0.079 & 268 & 0.79, 1.11 & 49 & 3,136 \\
     & (3.8 fm)$^3$ & & ($m_\pi L=5.0$) & 1.26 & 281 & 17,984 \\
     & & & & 1.42 & 294 & 28,224\\
\hline\hline
E5 & $64\times 32^3$ & 0.063 & 456 & 0.82 & 559 & 35,776 \\
     & (2.0 fm)$^3$ & & ($m_\pi L=4.7$) & 0.95 & 500 & 32,000\\
     & & & & 1.13 & 489 & 31,296\\
     & & & & 1.32 & 994 & 63,616\\
     & & & & 1.51 & 1,605 & 102,720 \\
\hline
F6 & $96\times 48^3$ & 0.063 & 324 & 0.82 & 60 & 3,840 \\
     & (3.0 fm)$^3$ & & ($m_\pi L=5.0$) & 0.95 & 150 & 9,600\\
     & & & & 1.07 & 75 & 4,800\\
     & & & & 1.32 & 254 & 16,256\\
     & & & & 1.20, 1.51 & 299 & 19,136\\
\hline
F7 & $96\times 48^3$ & 0.063 & 277 & 0.82, 0.95, 1.07 & 250 & 16,000 \\
     & (3.0 fm)$^3$ & & ($m_\pi L$=4.2) & 1.20, 1.32 & 250 & 32,000\\
     & & &  & 1.51 & 250 & 64,000\\
\hline
G8 & $128\times 64^3$ & 0.063 & 193 & 0.88 & 184 & 14,720 \\
     & (4.0 fm)$^3$ & & ($m_\pi L$=4.0) & 1.07 & 112 & 19,040\\
     &  & & & 1.26 & 182 & 29,120\\
     &  & & & 1.51 & 344 & 44,032\\
\hline\hline
N6 & $96\times 48^3$ & 0.05 & 332 & 0.9 & 110 & 3,520 \\
     & (2.4 fm)$^3$ & & ($m_\pi L$=4.1) & 1.1 & 888 & 28,416\\
     & & &  & 1.3, 1.5, 1.7 & 946 & 30,272\\
\hline\hline
\end{tabular}
\end{center}
\end{table}
\section{Numerical method}\label{sec:method}

\subsection{All-mode-averaging}

The all-mode-averaging (AMA) estimator
\cite{Blum:2012uh,Blum:2012my,Shintani:2014vja}
for an observable $\mathcal{O}$ can be defined as
\begin{eqnarray}
&&\mathcal O^{\rm AMA} = \mathcal O^{\rm (rest)} 
  + \frac{1}{N_G}\sum_{g\in G}^{N_G}\mathcal O^{{\rm (appx)}\,g},
\label{eq:caa1}\\
&&\mathcal O^{\rm (rest)} 
  = \frac{1}{N_{\rm org}}\sum_{f\in G}^{N_{\rm org}}\Big[\mathcal O^{f} - \mathcal O^{{\rm (appx)}\,f}\Big],
\label{eq:caa2}
\end{eqnarray}
where $\mathcal O^{\rm (appx)}$ denotes an approximate evaluation of
$\mathcal{O}$ constructed by means of applying a ``sloppy'' inversion
algorithm (a truncated solver with a precision of typically around $10^{-3}$)
to the Dirac operator. The bias inherent in any truncated-solver method
is corrected by the term $\mathcal{O}^{\rm (rest)}$. To ensure that the
expectation value of $\mathcal O^{\rm AMA}$ is consistent with $\mathcal O$,
both the sloppy and the exact evaluations of $\mathcal O$ are averaged
over orbits $\mathcal O^g$ under some subset $G$ (of size $N_G$) of a
symmetry group (such as translations) under which $\mathcal O$ transforms
covariantly.

To improve the accuracy with which the bias correction
$\mathcal{O}^{\rm (rest)}$ is estimated, it may also be averaged
over an orbit $\mathcal{O}^f$ under a subset of size
$N_{\rm org}\ll N_G$ of $G$. In this way, it becomes possible to reuse
existing exact evaluations of $\mathcal{O}$ using different
source positions to enhance the statistical accuracy without
having to recalculate the exact evaluations.

The idea behind AMA is that, as long as $\mathcal O^{\rm (appx)}$ is
an appropriate observable in the sense of having a strong correlation
with the original observable $\mathcal O$, the statistical accuracy
of $\mathcal O^{\rm AMA}$ evaluated on $N_{\rm conf}$ gauge configurations
should be similar to that of $\mathcal O$ on
$N_{\rm meas}=N_G\times N_{\rm conf}$ configurations, while the cost
of evaluating $\mathcal O^{\rm (appx)}$ is much lower than that of
evaluating $\mathcal O$. We should therefore expect to be able to
achieve a much-increased statistical accuracy for the same effort,
or conversely to have to pay only a reduced price for achieving a
desired statistical error.

More specifically, the ratio between the standard deviations of
$\mathcal O^{\rm AMA}$ and $\mathcal O$ is given by
\cite{Shintani:2014vja}
\begin{eqnarray}
\sigma^{\rm AMA}/\sigma &\simeq& \sqrt{ N_G^{-1} + 2\Delta r + R },\label{eq:ama1}\\
\Delta r &=& \Big(\frac{1}{N_{\rm org}}-\frac{1}{N_G}\Big)\Big(1-\frac{1}{N_{\rm org}}
\sum_{f\in G}^{N_{\rm org}}\frac{\langle\Delta\mathcal O^f\Delta\mathcal O^{{\rm (appx)}\,f}\rangle}
                         {\sigma^f\sigma^{{\rm (appx)}\,f}}\Big), \label{eq:ama2}\\
\quad R &=& \frac{1}{N^2_G}\sum_{g,g'\in G \atop g\not=g'}^{N_G}\frac{
    \langle\Delta\mathcal O^{{\rm (appx)}\,g}\Delta\mathcal O^{{\rm (appx)}\,g'}\rangle}
   {\sigma^{{\rm (appx)}\,g}\sigma^{{\rm (appx)}\,g'}}, \label{eq:ama3}
\end{eqnarray}
where $\Delta\mathcal O = \mathcal O - \langle\mathcal O\rangle$, 
and the standard deviation is given by
$\sigma = \sqrt{\langle(\Delta\mathcal O)^2\rangle}$.

The deviation from the ideal error-scaling behaviour
$\sigma^{\rm AMA}\sim 1/\sqrt{N_G}$ is parameterised by two quantities:
$\Delta r$ represents the degree of disagreement between $\mathcal O$
and $\mathcal O^{\rm (appx)}$ by tracking the amount by which the statistical
fluctuations of $\mathcal O^{\rm (appx)}$ fail to track those of $\mathcal O$,
while $R$ represents the amount by which using $N_G$ approximate measurements
$\mathcal O^{\rm (appx)\,g}$ falls short of providing $N_G$ statistically
independent measurements by tracking the degree to which the individual
approximations $\mathcal O^{{\rm (appx)}\,g}$ are correlated amongst
themselves. Note that we ignore the correlation between the exact measurements
$\mathcal O^f$ because these are generally sufficiently few in number
($N_{\rm org}\ll N_G$) that it is always possible to choose the spatial
separations between the sources large enough to render these correlations
negligible.

To reduce the error on $\mathcal O^{\rm AMA}$ as far as
possible, it is therefore desireable to achieve both $\Delta r\simeq 0$,
indicating close tracking of the exact by the approximate measurements,
and $R\simeq 0$, indicating nearly-independent measurements from
different source locations. Achieving the latter primarily relies on a
suitable choice of translation $g$ by large enough distances, whereas the
former has to be achieved by ensuring that the parameters of the truncated
solver used are suitably tuned.

\subsection{AMA with a locally deflated SAP+GCR solver}\label{subsec:sap}

So far, AMA has been mostly used with relatively inefficient solvers
such as CG. It is therefore worthwhile to study whether the significant
benefits reported in that context
\cite{Gupta:2016rli,Lin:2014saa}
carry over to the case of a more efficient solver, such as L\"uscher's
locally deflated SAP-preconditioned GCR solver
\cite{Luscher:2003qa,Luscher:2007se}
used in the DD-HMC
\cite{Luscher:2005rx,Luscher:2007es,DD-HMC},
MP-HMC
\cite{Marinkovic:2010eg},
and openQCD
\cite{Luscher:2012av,openQCD}
codes.

Here we recall the basic features of the Schwarz Alternating Procedure (SAP),
as discussed in refs.~\cite{Luscher:2003qa,Luscher:2007se}. When applied to
the Dirac equation
\begin{equation}
D\psi=\eta\,,
\end{equation}
the SAP is a ``divide and conquer'' strategy which starts by decomposing the
lattice into two non-overlapping domains $\Omega$ and $\Omega^*$ consisting
of blocks arranged in checkerboard fashion. The SAP then visits the blocks
in turn, updating the field on each block to the solution of the Dirac
equation with Dirichlet boundary conditions given by the field on the
neighbouring blocks. Due to the checkerboard structure of the block
decomposition, this can be done in parallel by simultaneously visiting
first all black blocks and then all white blocks in parallel.

Denoting the points of $\Omega$ and $\Omega^*$ that have neighbours
in $\Omega^*$ and $\Omega$, respectively, by $\partial\Omega^*$ and
$\partial\Omega$, the Dirac operator can be decomposed into a sum
\begin{equation}
D = D_\Omega + D_{\Omega^*} + D_{\partial\Omega} + D_{\partial\Omega^*}\,,
\end{equation}
where $D_\Omega$ acts only on the field at points $x\in\Omega$ with all
terms involving fields in $\Omega^*$ set to zero, $D_{\partial\Omega}$
contains the terms through which points in $\Omega$ receive contributions
from $\Omega^*$, and so forth. A complete cycle of the SAP can then be
written as
\begin{equation}
\psi \mapsto (1-KD)\psi + K\eta
\end{equation}
with the SAP kernel
\begin{equation}
K = D_\Omega^{-1} + D_{\Omega^*}^{-1} 
  - D_{\Omega^*}^{-1} D_{\partial\Omega^*} D_\Omega^{-1}.
\end{equation}
After $n_{\rm cy}$ SAP cycles starting from $\psi=0$, this corresponds
to approximating the inverse of the Dirac operator by the polynomial
\begin{equation}
D^{-1}\approx M_{\rm SAP} = K\sum_{\nu=0}^{n_{\rm cy}-1}(1-DK)^\nu\,,
\end{equation}
and this is used as a preconditioner by solving the right preconditioned
equation
\begin{equation}
D M_{\rm SAP} \phi = \eta
\end{equation}
using the generalised conjugate residual (GCR) algorithm
and setting $\psi = M_{\rm SAP} \phi$ at the end
\cite{Luscher:2003qa}.
The tunable parameters of the SAP preconditioner are therefore the
block size and the number $n_{\rm cy}$ of SAP cycles.

To further accelerate the GCR solver, deflation may be used as a means of
improving the condition number of the Dirac operator by separating the high
and low eigenmodes for separate treatment. If the deflation fields
$\{\phi_k\}_{k\le N}$ span a subspace (the deflation subspace)
containing good approximations to the low eigenmodes of the Dirac operator,
an oblique projector to the orthogonal complement of the deflation subspace
is given by
\begin{equation}
P_L = 1 - \sum_{k,l=1}^{N} D\phi_k \left(A^{-1}\right)_{kl} \phi_l^\dag \,,
\end{equation}
where
\begin{equation}
A_{kl} = (\phi_k,D\phi_l)
\label{eq:little}
\end{equation}
is called the little Dirac operator. The Dirac equation can then be split
into a low-mode and a high-mode part by left-projecting with $1-P_L$ and
$P_L$, respectively. The low-mode part can be solved in terms of the little
Dirac operator, so that the solution is given by
\begin{equation}
\psi = \chi + \sum_{k,l=1}^{N} \phi_k \left(A^{-1}\right)_{kl} (\phi_l,\eta)\,,
\label{eq:deflsoln}
\end{equation}
where the high-mode part $\chi=P_R\psi$ satisfies
\begin{equation}
P_L D \chi = P_L \eta\,, ~~~~ (1-P_R)\chi = 0
\end{equation}
with $P_L D = D P_R$
\cite{Luscher:2007se}.

Combining the block-decomposition approach with deflation leads to the
construction of a deflation subspace of dimension $N=N_s N_b$, where
$N_b$ is the number of blocks, and each deflation field has support only
on a single block. In practice, such deflation fields are obtained by
restricting a set of $N_s$ global deflation fields to each block and
orthonormalising the resulting fields using the Gram-Schmidt procedure.
To ensure that the deflation space approximates the low eigenspaces of
the Dirac operator efficiently, the global fields should ideally be good
approximations to the low modes. Such approximations can be obtained
using a few rounds of the inverse iteration $\phi_l \mapsto D^{-1}\phi_l$
starting from random fields. Since only approximate low modes are needed,
the exact inverse of the Dirac operator is not required, and the SAP
approximation $M_{\rm SAP}$ can be used instead
\cite{Luscher:2007se}.
Note that by using the block decomposition, we effectively are able to
obtain $N$ deflation vectors for the price of computing only $N_s\ll N$
approximate eigenmodes.
The tunable parameters of the deflation procedure are then the block size
and the number $N_s$ of global deflation fields.

For a more efficient implementation, mixed-precision calculations can be
used; since the SAP preconditioner needs not be very precise, single-precision
arithmetic suffices in this case. In the GCR algorithm, likewise, some
operations can be carried out in single precision \cite{Luscher:2003qa}.

In line with the setup used in the generation of the Monte Carlo ensembles,
we keep $n_{\rm cy}=5$ fixed in our setup.
The remaining algorithmic parameters that control the quality of
the sloppy solves, as measured by $\Delta r$ in Eq.(\ref{eq:ama2}) are then
the iteration number $N_{\rm iter}$ of the GCR algorithm,
the block size (which for simplicity we take to be the same
for the SAP preconditioner and the deflation procedure),
and the number $N_s$ of global deflation fields.
Tuning these parameters requires one to make a trade-off between the
quality of the AMA approximation and the gain in performance from using
the sloppy solver. In the case of the iteration number this is obvious,
while in the case of the deflation parameters the trade-off comes from
the increase of the size (and hence condition number) of the little Dirac
operator with $N_s$ and the number of blocks.
We will study the dependence of the overall runtime on these parameters
in section \ref{sec:performance} below.

A notable feature of the block-decomposition technique is that
for translations within the same block
the translation invariance of the approximate solution
$\mathcal O^{\rm (appx)}$ may be broken due to the limited
precision of the sloppy solver in conjunction with the Dirichlet
boundary conditions imposed in each SAP cycle.
To preserve the translational invariance of
$\mathcal O^{\rm (appx)\,g}$ under all transformations $g\in G$,
we use only shifts that map the domains $\Omega$ and $\Omega^*$
onto themselves.
As $M_{\rm SAP}$ is invariant under such translations,
these shifts are not affected by broken translation invariance.

\subsection{Computation of nucleon charges}

In this paper we concentrate on the application of AMA to computating the
axial ($g_A$), scalar ($g_S$) and  tensor ($g_T$) charges of the nucleon
in lattice QCD. 
These observables can be extracted from suitably renormalised ratios of
two- and three-point functions involving
the operators of the axial current $A_\mu=\bar\psi\gamma_\mu\gamma_5\psi$,
scalar density $S=\bar\psi\psi$, and
tensor current $T_{\mu\nu}=\bar\psi\sigma_{\mu\nu}\psi$, respectively.

For the nucleon, we use the interpolating field
\begin{equation}
  N_\alpha = \varepsilon^{abc}(u^{a} C\gamma_5 d^b)u_\alpha^c,  \label{eq:Nop}
\end{equation}
where $C$ is the charge conjugate matrix, $\alpha$ is the Dirac spinor index,
and $a,b,c$ are the color indices of the quark fields.
(In the following we will omit the spin indices from our notation).

In order to increase the overlap between the nucleon ground state and the
state created by applying the interpolating operator to the vacuum,
we apply Gaussian smearing
\cite{Gusken:1989ad}
(with APE-smeared
\cite{Albanese:1987ds}
gauge links in the Laplacian) at both source and sink. The smearing parameters
used are the same as in
\cite{Jager:2013kha,vonHippel:2013yfa}.

Using the spin-projection matrices $P^+=\frac{1}{2}(1+\gamma_0)$
and $P_{53}^+=P_+\gamma_5\gamma_3$, we evaluate the charges of the
nucleon through computing the ratios of three- and two-point functions
given
\begin{eqnarray}
  R_A^{\rm bare}(t,t_s) &=& \frac{{\rm tr}\Big[P_{53}^+\langle N(t_s) A_3(t)\bar N(0)\rangle\Big]}
                           {{\rm tr}\Big[P^+\langle N(t_s)\bar N(0)\rangle\Big]},\label{eq:R_A}\\
  R_S^{\rm bare}(t,t_s) &=& \frac{{\rm tr}\Big[P^+\langle N(t_s) S(t)\bar N(0)\rangle\Big]}
                           {{\rm tr}\Big[P^+\langle N(t_s)\bar N(0)\rangle\Big]},\label{eq:R_S}\\
  R_T^{\rm bare}(t,t_s) &=& \frac{{\rm tr}\Big[P_{53}^+\langle N(t_s) T_{12}(t)\bar N(0)\rangle\Big]}
                           {{\rm tr}\Big[P^+\langle N(t_s)\bar N(0)\rangle\Big]},\label{eq:R_T}
\end{eqnarray}
which yield the (bare) charges $g_A^{\rm bare}$, $g_S^{\rm bare}$,
and $g_T^{\rm bare}$, respectively, at asymptotically large time separations,
\begin{equation}
  \lim_{t_s,t_s-t\rightarrow \infty} R_{\mathcal O}^{\rm bare} 
     = g_{\mathcal O}^{\rm bare}.
\end{equation}
At finite time separations
the ratio $R_{\mathcal O}^{\rm bare}$ differs from its asymptotic value
$g_{\mathcal O}^{\rm bare}$
by time-dependent contributions from excited states with the same quantum
numbers as the ground state. We will discuss in section~\ref{sec:3pt} how
to best suppress these contributions, and how the use of AMA can help to
obtain a good signal even at relatively large time separations.

To further improve the statistical quality of the signal for the ratio
$R_{\mathcal O}^{\rm bare}$, we average over the forward- and
backward-propagating
nucleon, constructing the ratio separately for each direction in order to
take optimal advantage of correlations between the two- and three-point
functions. Obtaining the three-point function for both directions requires
the computation of sequential propagators with sink positions at both
$t_s$ and $T-t_s$, whereas the two-point function 
for both directions can be obtained
from a single inversion by using opposite parity projections for the forward
and backward directions.

We note that, since the charges are defined at zero momentum transfer
and we use the spatial components of the axial vector and tensor currents,
no additional operators are needed to realise O($a$) improvement.
To obtain the renormalised charges
$g_{\mathcal O}=Z_{\mathcal O}g_{\mathcal O}^{\rm bare}$,
we therefore only need to include the appropriate renormalisation constants,
which should ideally be chosen such that O($a$) improvement is realised.
We use the non-perturbative determination of $Z_A$ computed in the
Schr\"odinger functional
\cite{DellaMorte:2008xb}
together with the perturbative mass correction $b_A$ from
\cite{Sint:1997jx},
while for $Z_S$ and $Z_T$ we take the non-perturbative evaluations in
the $\rm \overline{MS}$ scheme at a renormalisation scale of $\mu=2$ GeV
using the RI-MOM scheme
\cite{Bali:2014nma}.
Since the values of the bare coupling used in
\cite{Bali:2014nma}
are slightly different from the ones used by us, we determine the
values of $Z_S$ and $Z_T$ to use by linear interpolation and
extrapolation in $1/\beta$, ignoring the unknown mass correction
terms.
In summary, our renormalised ratios are related to the bare ones by
\begin{align}
R_A(t,t_s) &= Z_A(1+b_A m_qa) R_A^{\rm bare}(t,t_s) \,, \label{eq:RAren}\\
R_S(t,t_s;\mu) &= Z_S(\mu) R_S^{\rm bare}(t,t_s) \,, \\
R_T(t,t_s;\mu) &= Z_T(\mu) R_T^{\rm bare}(t,t_s) \,.
\end{align}
Table~\ref{tab:zfact} shows the values of the renormalisation
constants used in this paper. 

\begin{table}
\begin{center}
\caption{Renormalisation factors for the axial current,
$Z_A^{\rm eff}=Z_A(1+b_A am_q)$, from \cite{DellaMorte:2008xb,Sint:1997jx},
and for
the scalar operator, $Z_S^{\rm \overline{MS}}(2~\rm{GeV})$,
and the tensor operator, $Z_S^{\rm \overline{MS}}(2~\rm{GeV})$,
from \cite{Bali:2014nma}. A perturbative error of the same order
as the one-loop contribution to $b_A$ is included in $Z_A^{\rm eff}$.}
\label{tab:zfact}
\begin{tabular}{cccccc}
\hline\hline
\\[-12pt] label & $Z_A^{\rm eff}$ & $Z_S^{\rm \overline{MS}}$ & $Z_T^{\rm \overline{MS}}$\\
\hline
A5 & 0.7785(83) & 0.6196(54) & 0.8356(15)\\
B6 & 0.7777(76) & 0.6196(54) & 0.8356(15)\\
E5 & 0.7866(83) & 0.6152(32) & 0.8540(91)\\
F6 & 0.7842(59) & 0.6152(32) & 0.8540(91)\\
F7 & 0.7835(51) & 0.6152(32) & 0.8540(91)\\
G8 & 0.7825(42) & 0.6152(32) & 0.8540(91)\\
N6 & 0.8022(91) & 0.6082(31) & 0.8886(95)\\
\hline\hline
\end{tabular}
\end{center}
\end{table}

\section{Tests and performance of AMA}\label{sec:performance}

In order to achieve the greatest possible gain in statistical accuracy
for fixed computational effort, we need to appropriately tune the parameters
of the sloppy solver.

\subsection{Covariance test}\label{subsec:test}

\begin{figure}[t]
\begin{center}
\includegraphics[width=90mm]{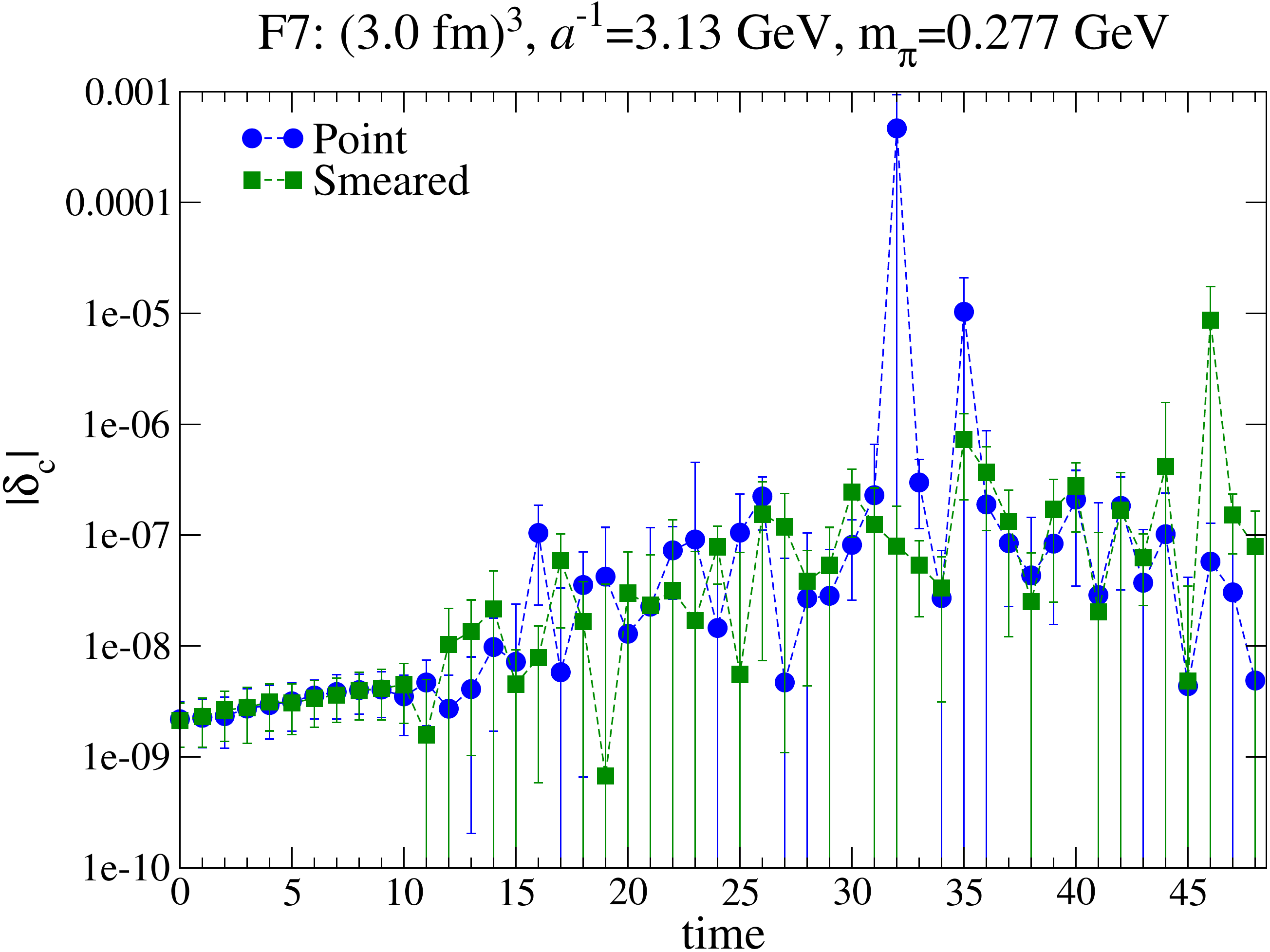}
\caption{The covariance violation $|\delta_c|$ defined in (\ref{eq:delta_c})
for the nucleon two-point function as a function of source-sink separation
as measured on one F7 configuration ($96\times48^3$ lattice, $m_\pi=0.277$ GeV).
The different colors indicate the results obtained using smeared (green squares)
and point (blue circles) sources and sinks, respectively.}\label{fig:test}
\end{center}
\end{figure}

Before discussing how to tune the parameters of AMA for use with
the deflated SAP-preconditioned GCR solver, we first need to check
that the covariant symmetry of the approximation is preserved in the
presence of the domain decomposition underlying the SAP preconditioner.

The amount by which the assumed covariance is violated can be parameterized
by \cite{Shintani:2014vja}
\begin{equation}
 \delta_c = (\mathcal O^{{\rm (appx)}\,g}[U^{\bar g}] - \mathcal O^{(\rm
appx)}[U])/\mathcal O^{(\rm appx)}[U],
 \label{eq:delta_c}
\end{equation}
i.e. the relative difference between evaluating the original observable on
the original gauge field and evaluating the approximation based on the
transformation $g$ on the appropriately transformed gauge field; if
covariance were exact, we would find $\delta_c=0$.

For a numerical test, we use the F7 ensemble from Table~\ref{tab:latparam},
choosing a domain size of $6^4$ for the SAP preconditioner. 
In line with the arguments laid out in
Section \ref{subsec:sap}, shift vectors need be such that each of
the SAP subdomains is mapped onto itself; we therefore choose the shifted
source location to be $g=(6,6,0,0)$, with the corresponding shift of the gauge
field given by $\bar g=(-6,-6,0,0)$.
Since the point of this comparison is to check that the effects of the block
decomposition in the SAP are under control, we did not use deflation for this
test. For the approximation, we therefore had to choose a fixed iteration
count of $N_{\rm iter}=30$
for the GCR algorithm, which in our example corresponds to a residual norm
of order $10^{-2}$.
The resulting covariance violation $|\delta_c|$ for the nucleon two-point
function as a function of the source-sink separation is shown in
Figure~\ref{fig:test}.
At large time separations $t\gtrsim 30a$, the accumulation of round-off errors
in the approximation leads to a non-negligible covariance violation
$|\delta_c|\sim 10^{-5}-10^{-3}$ (which however is still less than the
statistical errors in this time region). On the other hand, the covariance
assumption is well-justified in the typical signal region $t\lesssim 25a$,
where $|\delta_c|\lesssim 10^{-6}$. We may therefore conclude that with this
set of parameters, the systematic error arising from violating the covariance
assumption is negligible for practical purposes.

\subsection{Correlation between original and approximation}

\begin{figure}
\begin{center}
\vskip 10mm
\includegraphics[width=73mm]{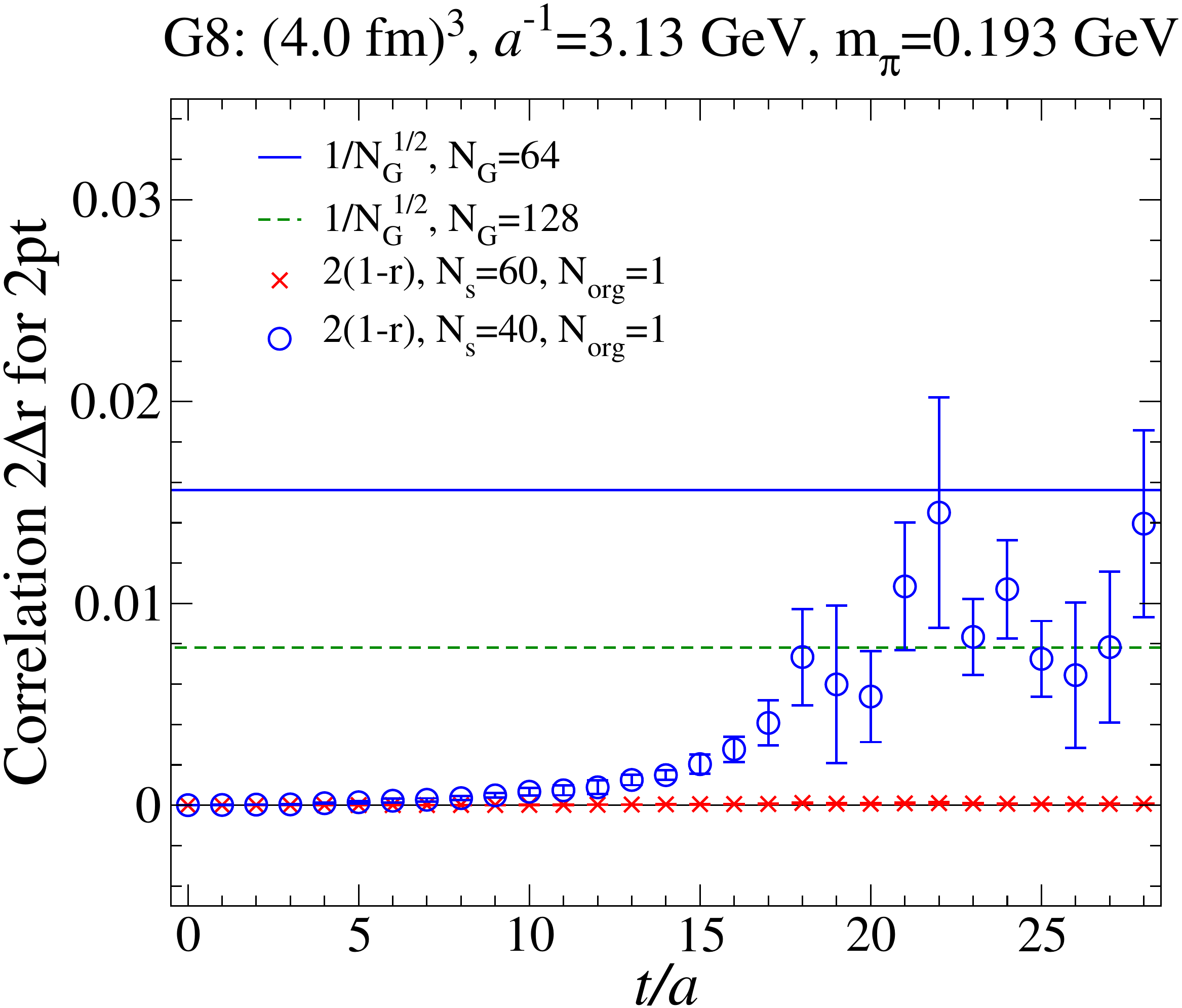}
\hspace{2mm}
\includegraphics[width=80mm]{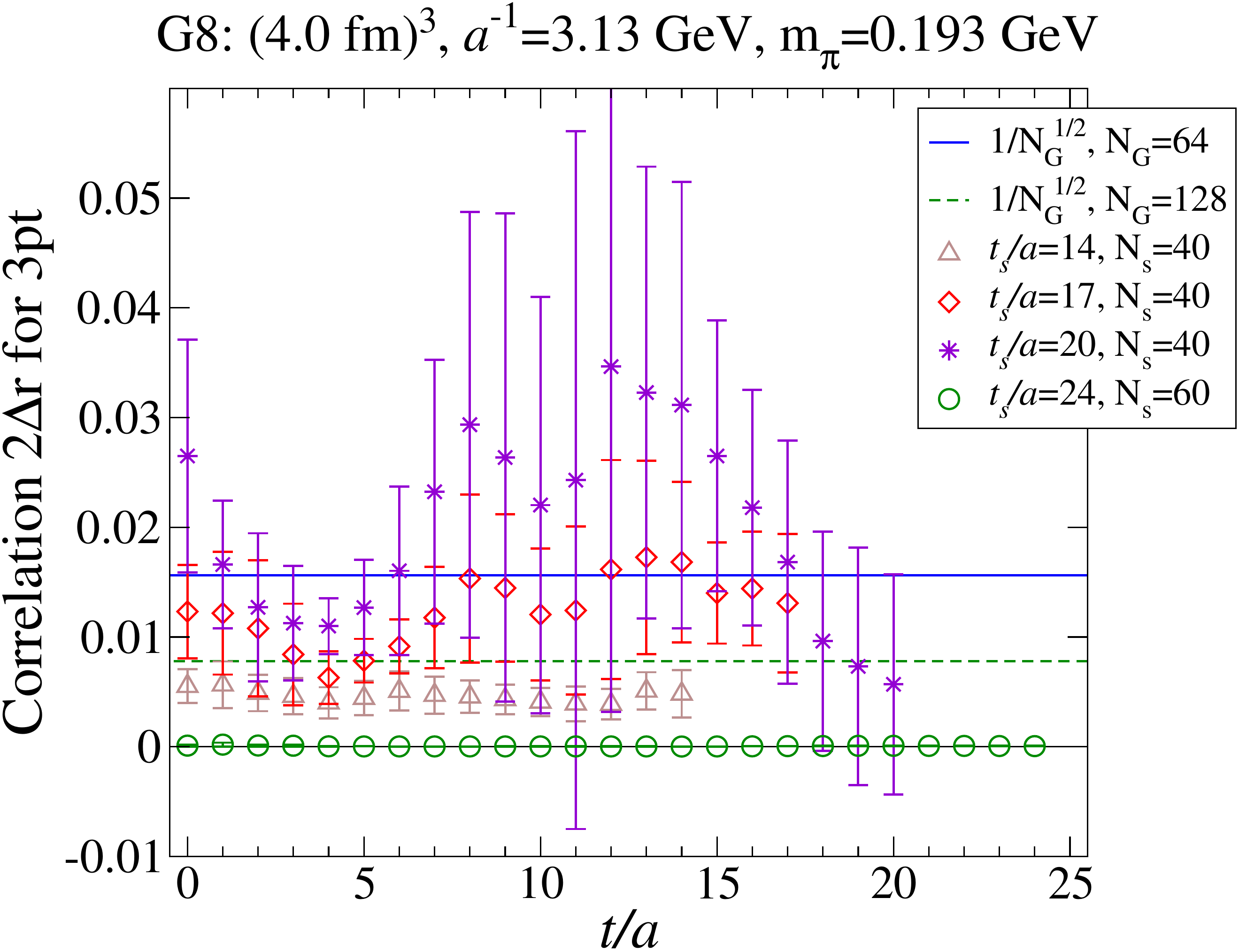}
\caption{The correlation mismatch $2\Delta r$ on the G8 ensemble
($128\times 64^3$ lattice, $m_\pi=0.193$~GeV)
for the nucleon two-point function (left)
and the axial-vector nucleon three-point function (right),
as a function of the source-sink separation $t_s$ and operator insertion
time $t$.
The straight lines indicate the level of the statistical error reduction
achievable with AMA in the case of mutually perfectly decorrelated
approximations that are each perfectly correlated with the original
observable.}
\label{fig:r_N_ga}
\end{center}
\end{figure}

Given that the statistical error of the AMA result depends crucially on the
discrepancy $\Delta r$ between the fluctuations of the original observable
and its approximate evaluation, our tuning of the solver parameters will have
to be guided by considering the parameter dependence of $\Delta r$.

The left panel of Figure~\ref{fig:r_N_ga} shows $2\Delta r$ for the nucleon
two-point function as a function of source-sink separation using both
$N_s=40$ and $N_s=60$ deflation vectors. For $N_s=40$, an increase in
$\Delta r$ can be seen at larger time separations $15\le t/a\le 25$,
where it becomes large enough to no longer be negligible compared to
$1/N_G$ for $N_G=128$.
For $N_s=60$, on the other hand, we find $\Delta r\lesssim 10^{-3}$
all the way out to $t/a=25$ (corresponding to a separation of $\sim 1.5$~fm).
The optimal choice of the parameters for the approximation therefore will
in general depend on the maximal time separation one is interested in.

The right panel of Figure~\ref{fig:r_N_ga} shows $\Delta r$ for the
three-point function appearing in the numerator of the ratio $R_A$
as a function of the operator insertion time $t$ for a range of different
source-sink separations $t_s$. For $N_s=40$, $\Delta r$ is seen to increase
with $t_s$, becoming comparable to $1/N_G$ for $N_G=128$ at $t_s/a=20$.
Since $\Delta r$ can be reduced by increasing $N_{\rm org}$ as per
Eq.~(\ref{eq:ama2}), we use $N_{\rm org}>1$  for $t_s/a=20$.
For $N_s=60$, we see that $\Delta r$ is sufficiently reduced to be
negligible even at $t_s/a=24$, albeit at the expense of a 1.6-fold increase
in computing time (cf.~Figure~\ref{fig:rate_ga}),
indicating that the trade-off between computational cost
and achievable statistical accuracy is a crucial consideration
in tuning the parameters of the approximation used in AMA.

\subsection{Performance of AMA for different approximation parameters}\label{subsec:performance}

\begin{figure}
\begin{center}
\includegraphics[width=85mm]{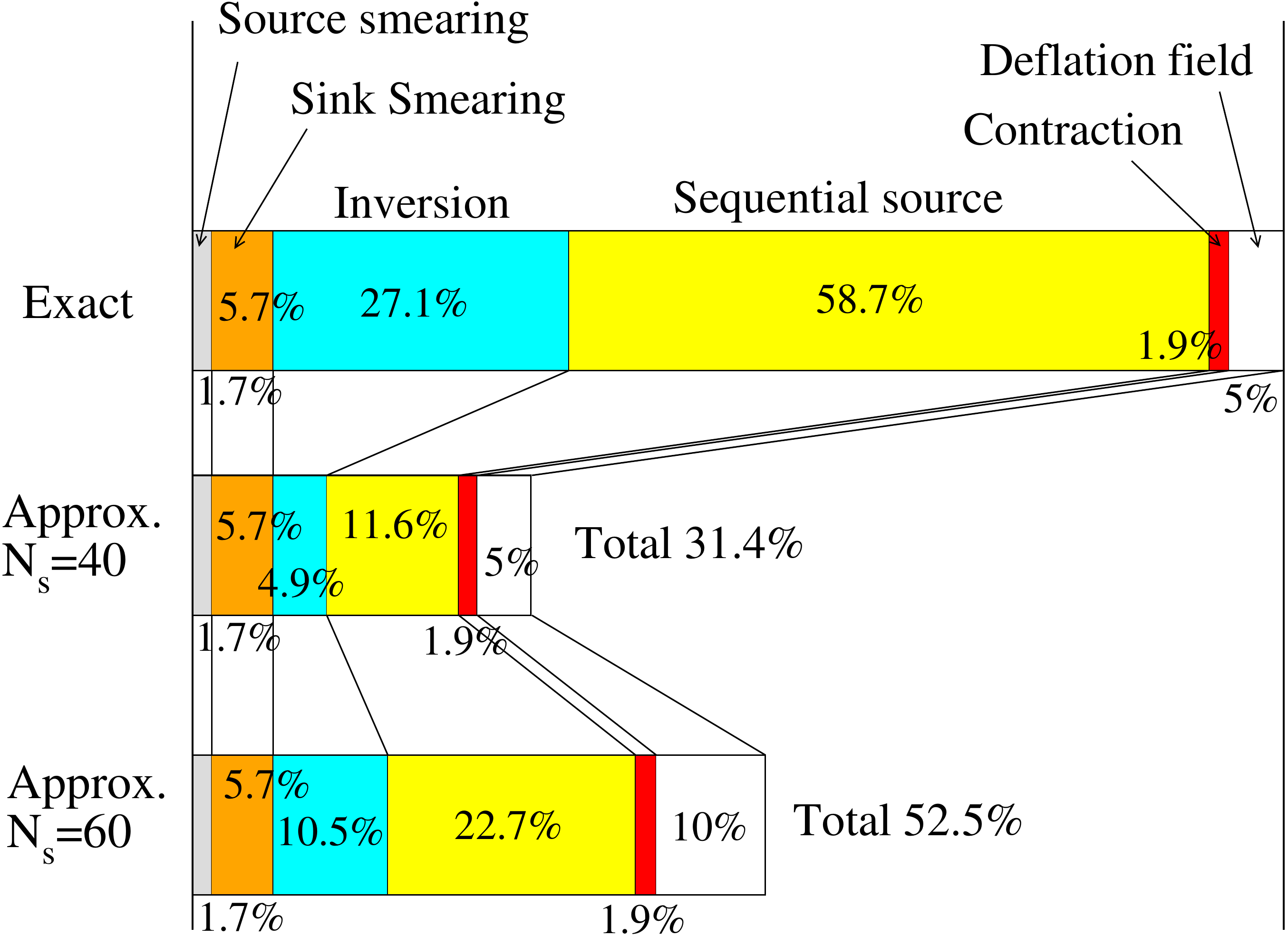}
\caption{A comparison of the computational cost between the exact
(using $N_s=40$ deflation vectors)
and two approximate evaluations (using $N_s=40$ and $N_s=60$, respectively)
of the axial charge of the nucleon on the G8 ensemble. All percentages
given are relative to the total time for the exact evaluation.}
\label{fig:rate_ga}
\end{center}
\end{figure}

In Figure \ref{fig:rate_ga}, we show a comparison of the overall performance
of two different approximations (using $N_s=40$ and $N_s=60$ deflation
vectors, respectively) relative to the exact evaluation (with $N_s=40$
deflation vectors), using the G8 ensemble as a test case.
For the approximation with $N_s=40$, the time required for inverting
the Dirac operator is reduced by a factor of 5, whereas for $N_s=60$,
the reduction is only by a factor of 3 due to the larger size of the ``little''
Dirac operator (\ref{eq:little}) in Eq.~(\ref{eq:deflsoln}).
Taking into account the fact that generating a larger number of deflation
fields is also more costly, and including the fixed costs of source and
sink smearing and contractions, the total time for an approximate
evaluation using $N_s=40$ is about 30\% of the exact calculation,
whereas for $N_s=60$ it is about 50\%.

\subsection{Error scaling and computational cost}

\begin{figure}[t]
\begin{center}
\includegraphics[width=80mm]{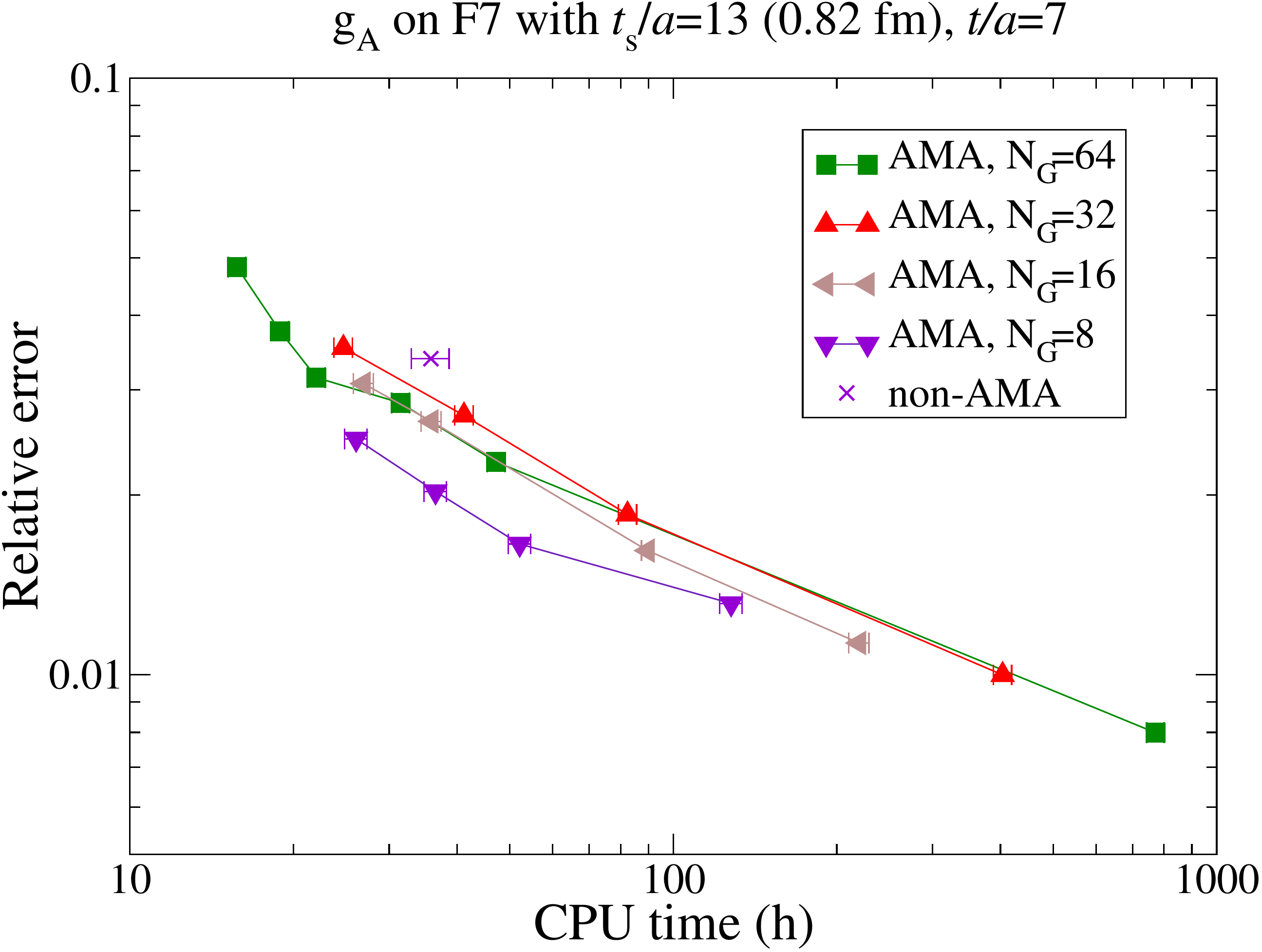}
\vskip 3mm
\caption{The relative error of the axial charge measured on the
F7 ensemble using $t_s/a=13$, $t/a=7$ plotted against the total CPU time
used.
The different symbols and colours denote different values for $N_G$,
where the error scaling for each is explored using different numbers
of gauge configurations.
In the case of the non-AMA evaluation, the complete gauge ensemble is used
with a single source position for the exact evaluation.}
\label{fig:cputime}
\end{center}
\end{figure}

Finally, we consider how the overall statistical error scales as a function
of the total CPU time used when varying the total number of measurements by
varying $N_G$, $N_{\rm conf}$, or both.

Figure \ref{fig:cputime} shows the relative error of the ratio $R_A(t,t_s)$
from Eq.~(\ref{eq:RAren}), as measured on the F7 ensemble using $t_s/a=13$ and
$t/a=7$, plotted against the total CPU time used for AMA with
$N_G=8$, $16$, $32$, and $64$, with different numbers of configurations used.
The relative error and CPU time required when not using AMA and performing
a single exact evaluation on each configuration instead is also shown for
comparison.

Since increasing $N_G$ leads to an increase in the correlation $R$ between
the different approximations (because the source positions $g$ will have to
be taken closer together when using more approximate solves), the
error scaling between different values of $N_G$ is not perfect.
In our test case, using $N_G=8$ requires about half the computational cost at
the same relative error as the larger values of $N_G$ considered, whereas
the larger $N_G$ values exhibited roughly identical error scaling.

Comparing AMA to the conventional method, a reduction in error by a factor
of about $1.5-3$ can be achieved at constant computational cost by using AMA.
This is one of the main findings of this paper concerning the performance
and effectiveness of AMA in conjunction with highly efficient solvers.

\subsection{Tuning of AMA parameters}\label{subsec:tuning}

\begin{table}[t]
\begin{center}
\caption{AMA tuning parameters for each gauge ensemble.}
\label{tab:param}
\begin{tabular}{ccccccccc}
\hline\hline
Label & Domain size & $N_s$ & $N_G$ & $N_{\rm iter}$ & $N_{\rm org}$ & $t_s$ [fm]  \\
& (T$\times$ X$\times$ Y$\times$ Z) & & & (2pt, 3pt) \\
\hline\hline
A5 & 4$\times$4$\times$4$\times$4 & 30 & 64 & (3,3) & 1& 0.79, 0.95, 1.11\\
     & & & & & & 1.26, 1.42 \\\cline{3-7}
\hline
B6 & 6$\times$6$\times$6$\times$6 & 40 & 64 & (4,3) & 1& 0.79, 1.11, 1.26\\\cline{3-7}
     & & 40 & 96 & (3,3) & 1& 1.42\\\cline{3-7}
\hline\hline
E5 & 4$\times$4$\times$8$\times$8 & 30 & 64 & (3,3) & 1& 0.82, 0.95,\\
     & & & & & & 1.13, 1.32 \\\cline{3-7}
     &  & 30(994 cfgs.) 
             & \multirow{2}{*}{64} & \multirow{2}{*}{(3,3)} & \multirow{2}{*}{5} & \multirow{2}{*}{1.51}\\
     &  & 40(661 cfgs.)\\\cline{3-7}
\hline
F6 & 6$\times$6$\times$6$\times$6 & 30 & 64 & (4,3)& 1& 0.82, 0.95, 1.07\\
     & & & & & & 1.20, 1.32 \\\cline{3-7}
     & & 40 & 64 & (4,3) & 1& 1.51\\\cline{3-7}
\hline
F7 & 6$\times$6$\times$6$\times$6 & 30 & 64 & (4,3)& 1& 0.82, 0.95, 1.07\\\cline{3-7}
     &  & 30 & 64 & \multirow{2}{*}{(4,3)} & \multirow{2}{*}{4} & \multirow{2}{*}{1.2, 1.32}\\
     &  & 40 & 64 \\\cline{3-7}
     &  & 30 & 192 & \multirow{2}{*}{(4,3)} & \multirow{2}{*}{15} & \multirow{2}{*}{1.51}\\
     &  & 40 & 64 \\\cline{3-7}
\hline
G8 & 8$\times$8$\times$8$\times$4 & 40 & 80 & (4,3) & 1& 0.88\\\cline{3-7}
     &  & 40 & 170 & (4,3) & 1& 1.07\\\cline{3-7}
     &  & 40(101 cfgs.) & 160 & \multirow{2}{*}{(4,3)} & \multirow{2}{*}{5} & \multirow{2}{*}{1.26}\\
     &  & 50(81 cfgs.) & 160 \\\cline{3-7}
     &  & 60 & 128 & (4,3) & 1 & 1.51\\\cline{3-7}
\hline\hline
N6 & 6$\times$6$\times$6$\times$6 & 30 & 64 & (4,3) & 1& 0.9, 1.1, 1.3\\
     & & & & & & 1.5, 1.7\\\cline{3-7}
\hline\hline
\end{tabular}
\end{center}
\end{table}

For the remaining calculations, we have tuned the parameters of the
deflated SAP-preconditioned GCR solver so as to achieve the maximum
reduction in computational cost while keeping $\Delta r$ sufficiently
suppressed to enable good error scaling.
Table \ref{tab:param} shows the resulting tuned parameter values.

We use fixed numbers $N_{\rm iter}$ of GCR iterations
for the point-to-all propagator and the sequential propagator, as shown
in Table~\ref{tab:param};
using a sloppier propagator between the sink and operator insertion point
does not lead to an increase of $\Delta r$, and thus enables us to reduce the 
computational cost for the three-point function.
We note that the iteration count
of $N_{\rm iter}=3$ corresponds to a similar residual as $N_{\rm iter}=30$
in the undeflated test of section~\ref{subsec:test}.

\section{AMA study of excited-state contamination in nucleon two-point functions}
\label{sec:2pt}

\begin{figure}
\begin{center}
\includegraphics[width=72mm]{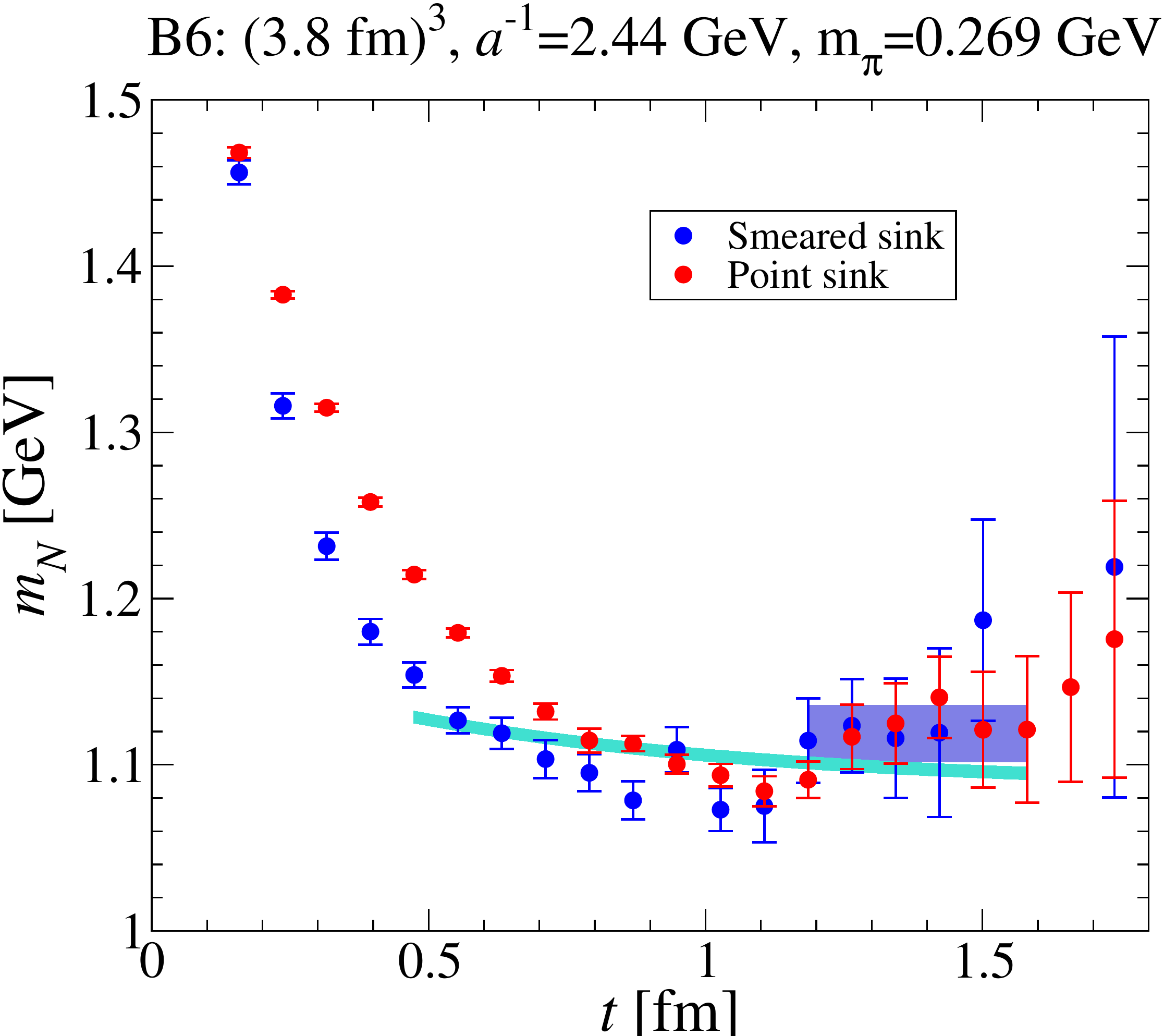}
\hspace{5mm}
\includegraphics[width=72mm]{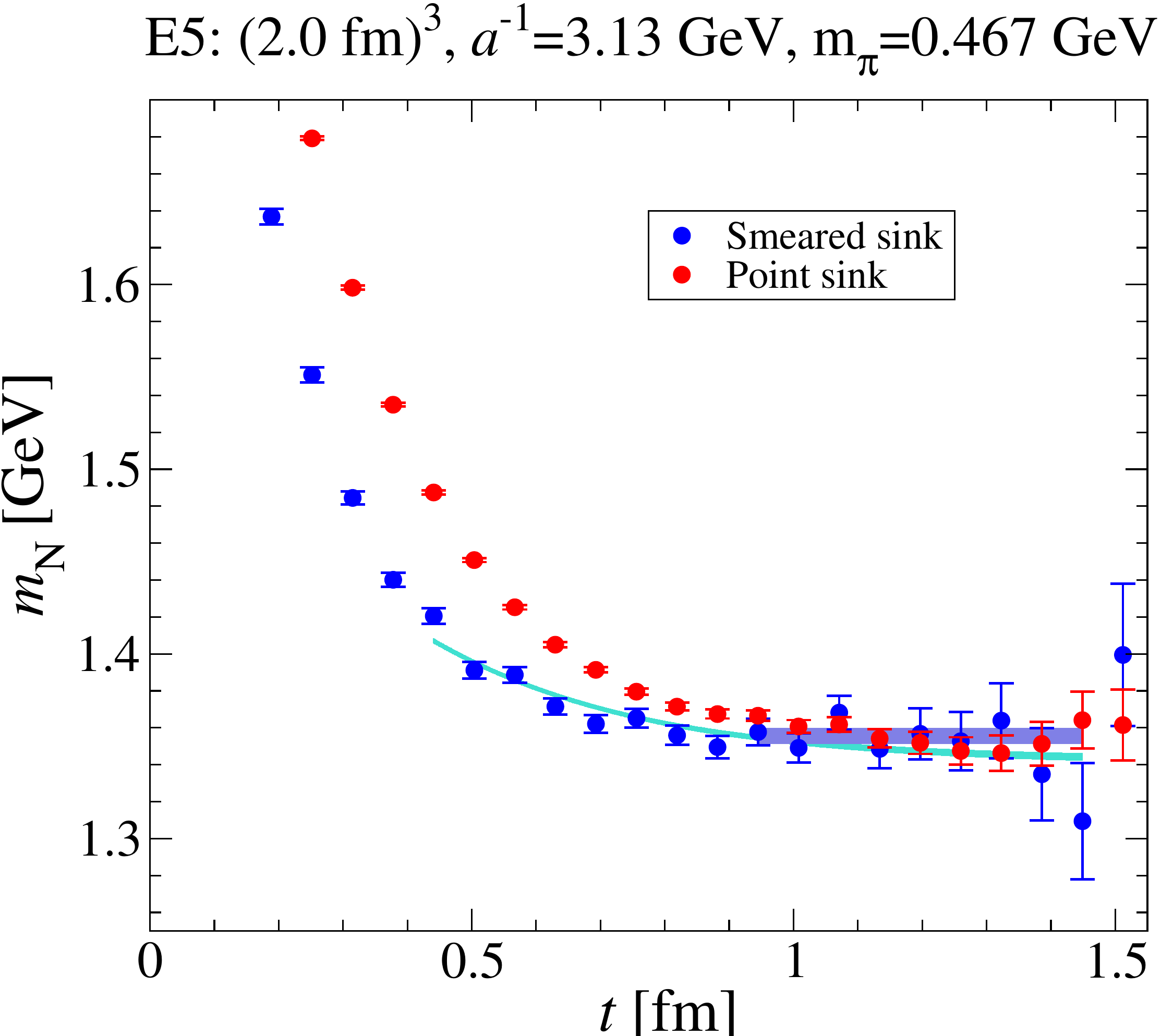}
\vskip 5mm
\includegraphics[width=72mm]{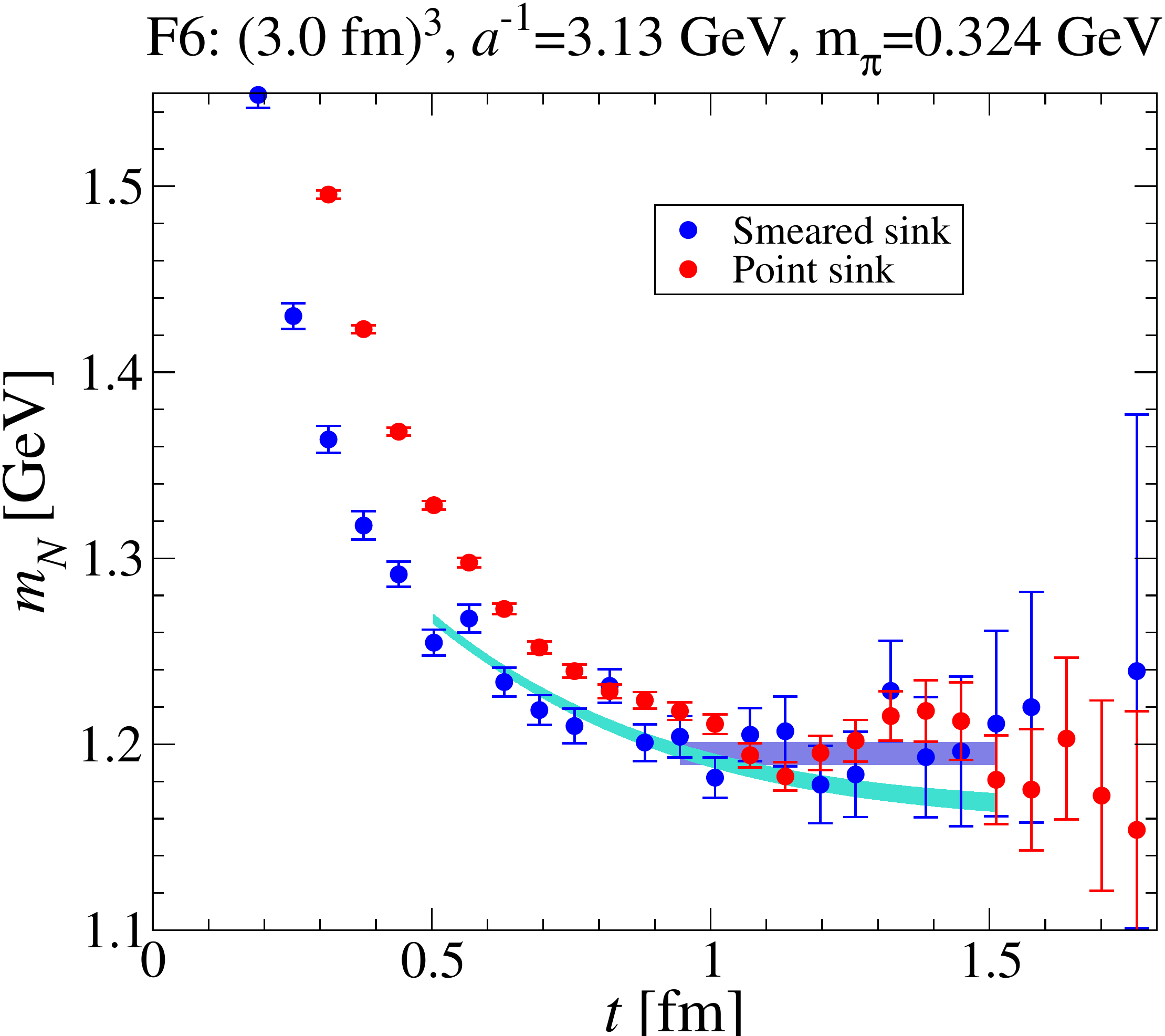}
\hspace{5mm}
\includegraphics[width=72mm]{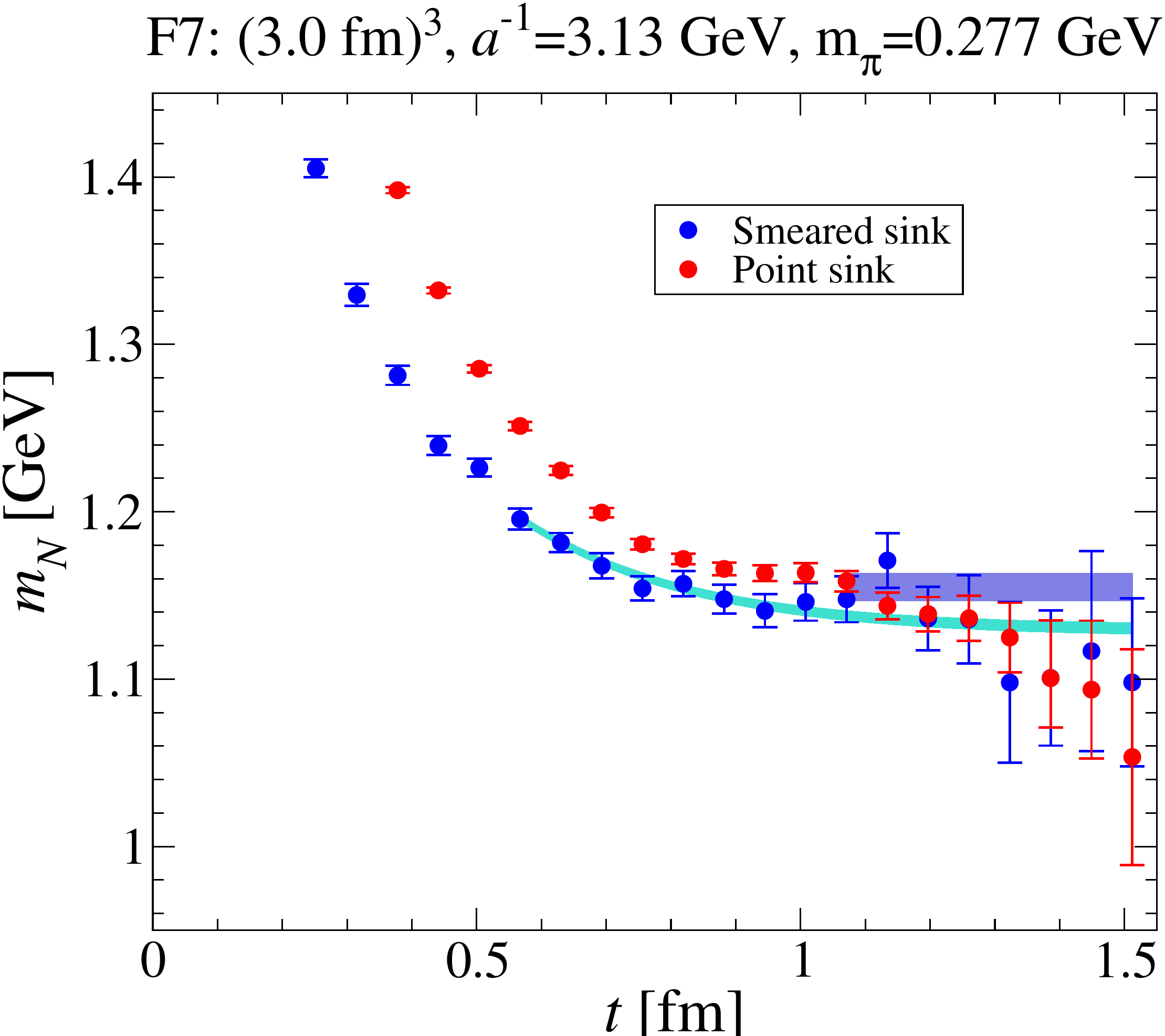}
\vskip 5mm
\includegraphics[width=72mm]{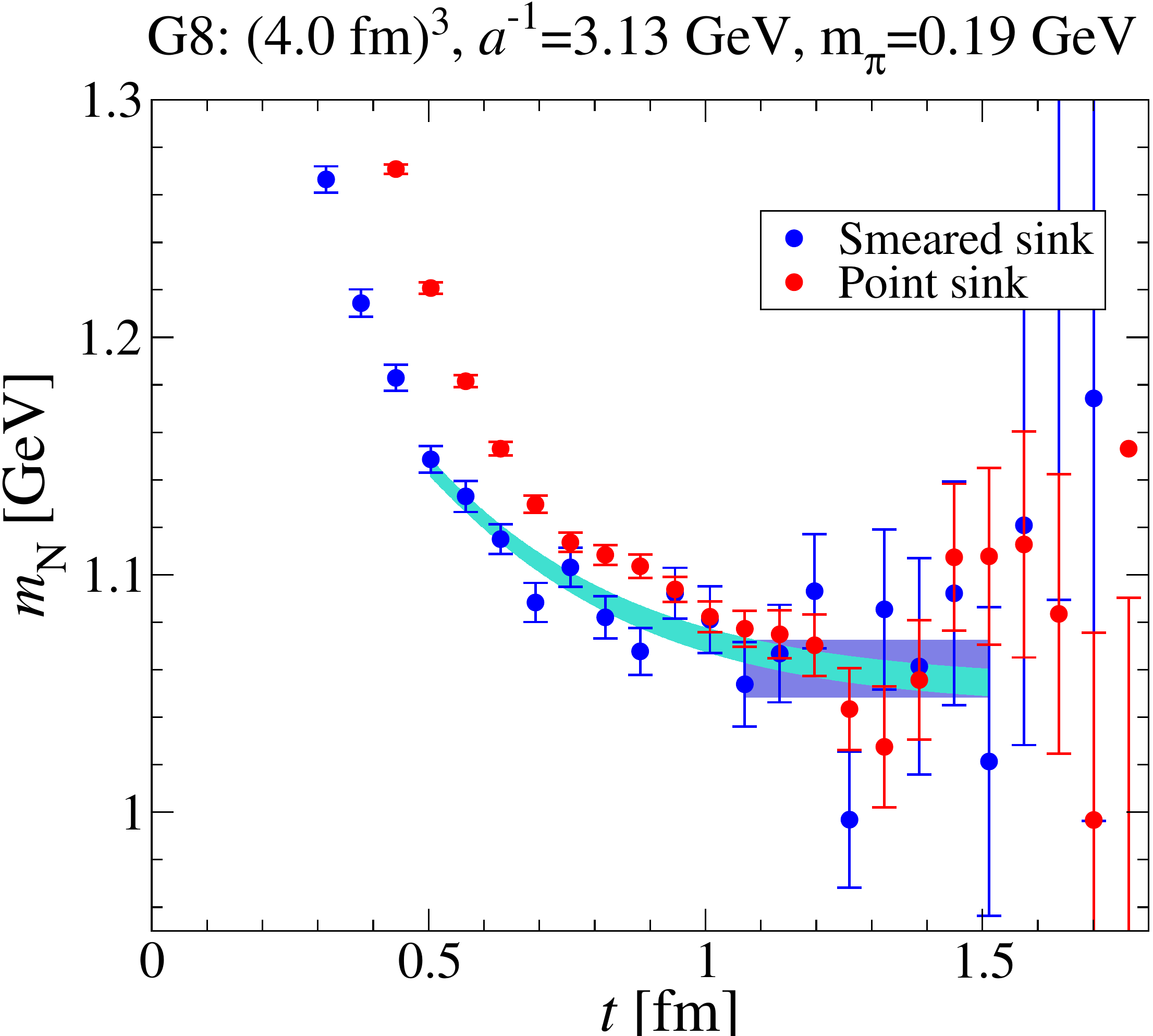}
\hspace{5mm}
\includegraphics[width=72mm]{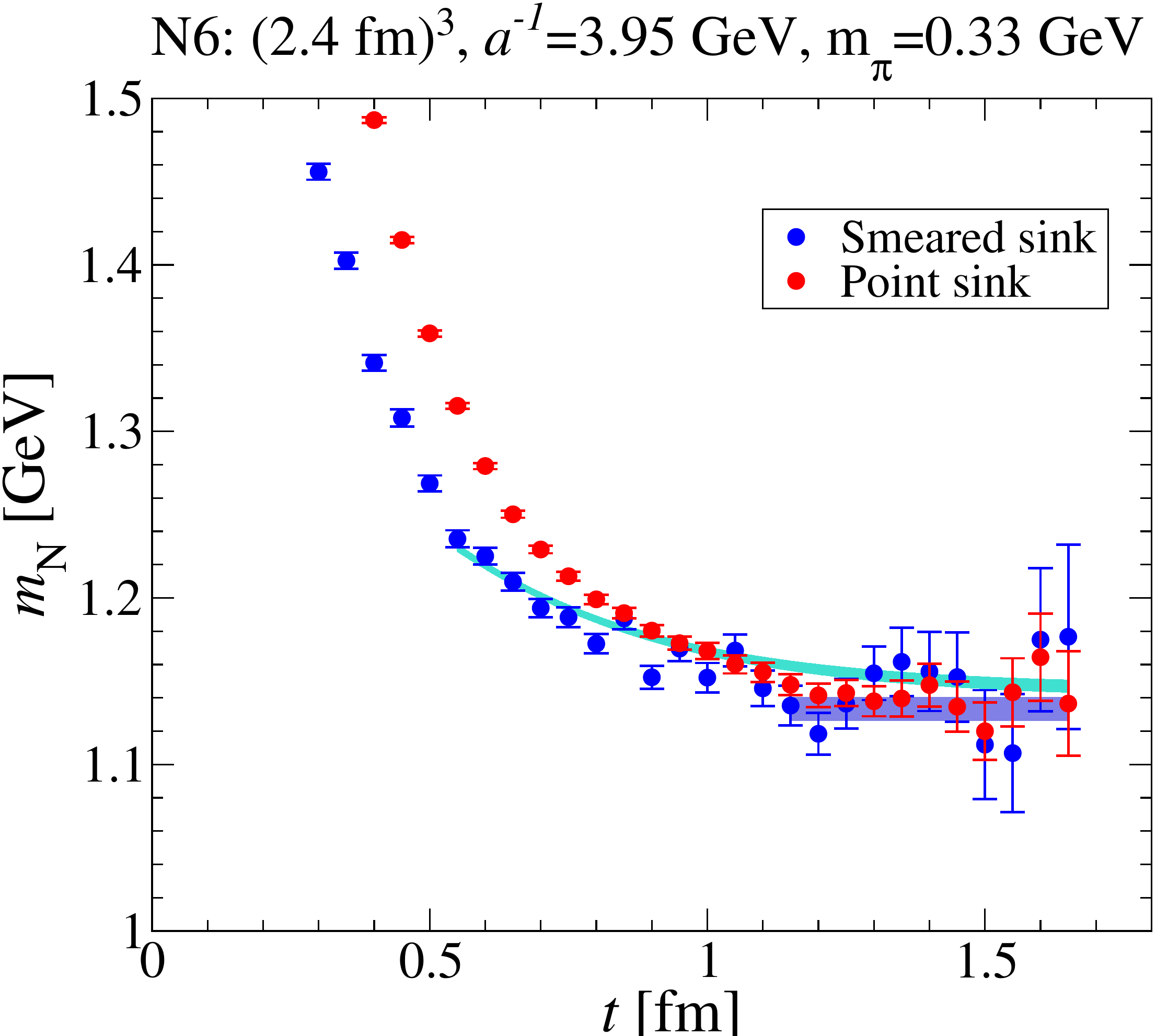}
\caption{Effective mass plots of the nucleon for six gauge ensembles.
The ensemble parameters are indicated in the plots. The two colours 
indicate the different nucleon sink operators. The blue and 
cyan bands indicate the results and errors from single- and double-exponential 
fits to the correlator.}
\label{fig:effm}
\end{center}
\end{figure}

The nucleon two-point correlation function may be approximated as 
\begin{eqnarray}
  C_N(t) \simeq Z_N e^{-m_Nt} + Z_{N'}e^{-m_{N'}t}+\cdots,
  \label{eq:fit_nn}
\end{eqnarray}
with masses $m_{N}$ and $m_{N'}$ and overlap factors $Z_{N}$ and $Z_{N'}$,
for the ground state and first excited state respectively. In order to study the 
nucleon mass, we utilised single- and double-exponential fits to the correlation function 
eq.~\eqref{eq:fit_nn}, of the form
\begin{eqnarray}
  C_N(t) &=& A e^{-mt}, \\
  C_N(t) &=& A e^{-mt} + B e^{-(m+\Delta)t}.
  \label{eq:ans_nn}
\end{eqnarray}
 For the fitting, we used a $\chi$-squared minimisation and found that the nucleon mass
 could be determined reliably using a single-exponential fit performed in the interval
 1.0~fm$~< t<~$1.5~fm (confirmed by the double-exponential fits), whereas in order to 
incorporate the excited states fitted in the double-exponential fits, an earlier fitting interval 
was required, starting at $t\simeq 0.5$ fm. This is highlighted by the 
plots in Figure~\ref{fig:effm}, where the effective mass
\begin{equation}
  m^{\rm eff}_N(t) = \ln \frac{C_N(t)}{C_N(t+1)}, 
\end{equation}
is used to monitor the region of ground state dominance and of
excited-state contamination. It also indicates the effectiveness of the 
exponential fits.  

\begin{figure}
\begin{center}
\includegraphics[width=120mm]{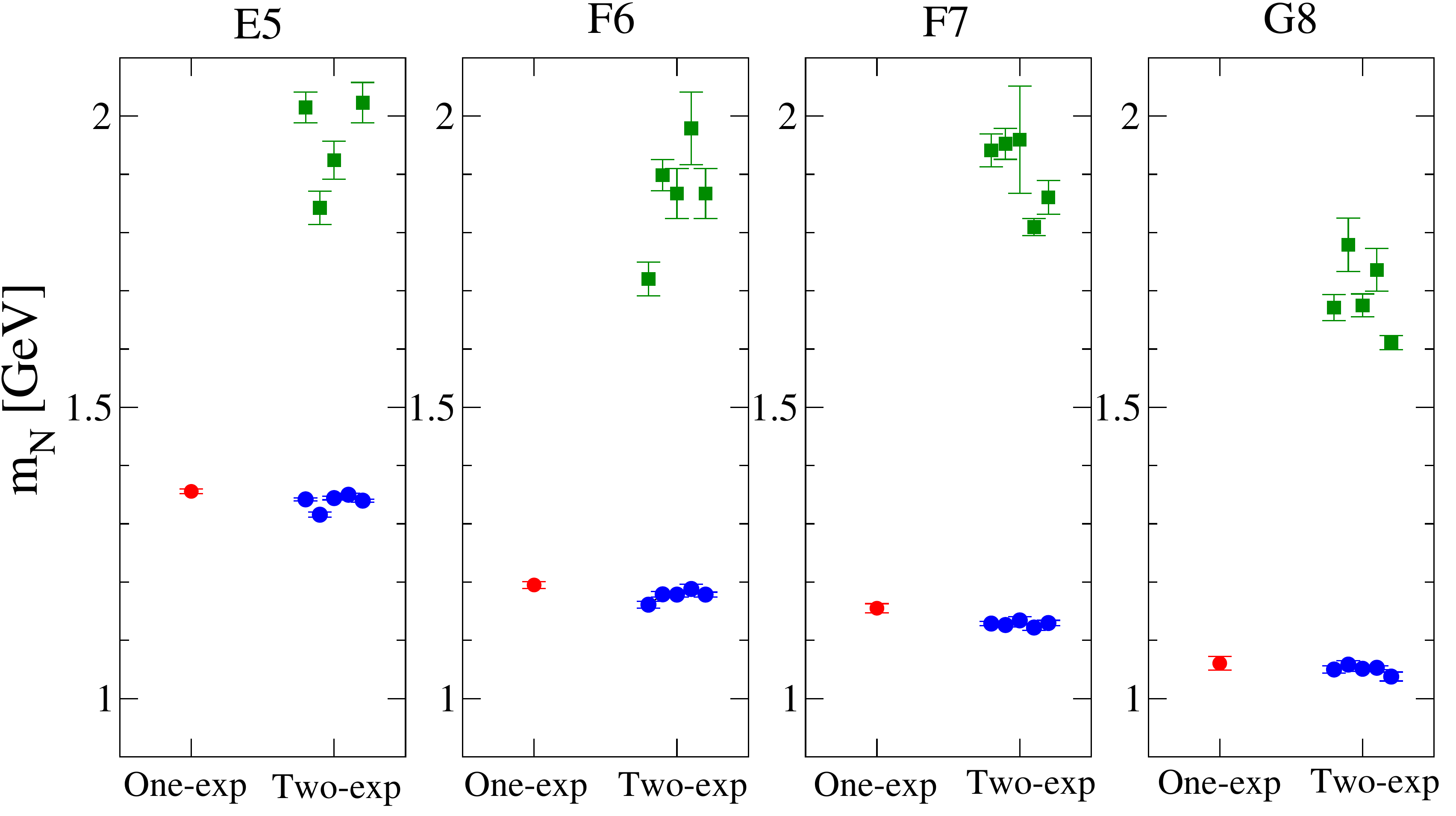}
\caption{Results for single- (red points) and double-exponential fits to the nucleon two-point function for the $\beta=5.3$ gauge ensembles. For the double-exponential fits, the blue and green points
indicate the ground and excited state masses respectively, for a number of
different fitting intervals (indicated in Table \ref{tab:fit_2pt}).}
\label{fig:fexp2}
\end{center}
\end{figure}

\begin{table}
\begin{center}
\caption{Results for single- and double-exponential fits to the nucleon two-point function 
on each gauge ensemble. The nucleon mass, $\chi^2$/dof and fitting 
intervals (in lattice units) are given. To determine the masses, the largest set of measurements
available in Table \ref{tab:latparam} for each ensemble was used (i.e. for ensemble E5 the 
measurement set where $N_{\mathrm{meas}}=102,720$ was used)}
\label{tab:fit_2pt}
\begin{tabular}{c|c|ccccc|c}
\hline\hline
   & Single-exp. & Double-exp. & & & & & Ref.~\cite{Capitani:2015sba}\\
\hline\hline
A5 & [15,20] & [6,20] & [5,20] & [7,20] & [6,19] & [6,21] & [10,25]\\
$am_N$ & 0.465(6) & 0.437(1) & 0.444(2) & 0.444(2) & 0.437(2) & 0.449(1) & 0.468(7)\\
$\chi^2$/dof & 0.8 & 4.6 & 1.54 & 2.9 & 2.6 & 4.8 & 0.9\\
\hline
B6 & [15,20] & [6,20] & [5,20] & [7,20] & [6,19] & [6,21] & [8,20]\\
$am_N$ & 0.448(7) & 0.436(2) & 0.444(2) & 0.424(1) & 0.433(4) & 0.434(2) & 0.444(5)\\
$\chi^2$/dof & 0.4 & 3.8 & 13.2 & 2.3 & 1.0 & 1.2 & 1.3\\
\hline\hline
E5 & [15,23] & [7,23] & [6,23] & [8,23] & [7,22] & [7,24] & [11,25]\\
$a m_N$ & 0.433(1) & 0.429(1) & 0.421(1) & 0.430(1) & 0.432(1) & 0.429(2) & 0.441(4)\\
$\chi^2$/dof & 0.5 & 2.5 & 5.3 & 1.6 & 0.6 & 2.2 & 1.1\\
\hline
F6 & [15,24] & [8,24] & [7,24] & [9,24] & [8,23] & [8,25] & [11,25]\\
$am_N$ & 0.382(2) & 0.370(2) & 0.376(1) & 0.376(1) & 0.379(3) & 0.376(1) & 0.382(4)\\
$\chi^2$/dof & 0.6 & 1.9 & 1.5 & 1.6 & 1.6 & 1.6 & 1.0\\
\hline
F7 & [17,24] & [9,24] & [8,24] & [10,24] & [9,23] & [9,25] & [11,25]\\
$a m_N$ & 0.369(3) & 0.361(1) & 0.360(1) & 0.363(2) & 0.359(2) & 0.361(2) & 0.367(5)\\
$\chi^2$/dof & 0.9 & 1.1 & 1.3 & 1.1 & 1.5 & 1.2 & 0.58\\
\hline
G8 & [17,24] & [8,24] & [7,24] & [9,24] & [8,23] & [8,25] & [11,24]\\
$am_N$ & 0.338(4) & 0.335(2) & 0.338(2) & 0.336(1) & 0.336(1) & 0.331(2) & 0.352(6)\\
$\chi^2$/dof & 1.3 & 1.5 & 1.8 & 0.9 & 1.6 & 1.6 & 1.4\\
\hline\hline
N6 & [23,33] & [11,33] & [10,33] & [12,33] & [11,32] & [11,34] & [15,30]\\
$a m_N$  & 0.288(2) & 0.290(1) & 0.285(2) & 0.290(7) & 0.293(1) & 0.290(1) & 0.297(3)\\
$\chi^2$/dof & 1.3 & 2.7 & 2.2 & 3.3 & 2.7 & 2.6 & 0.69\\
\hline\hline
\end{tabular}
\end{center}
\end{table}

The results for the single- and double-exponential fits to the correlation function for each of the  ensembles are given in Table \ref{tab:fit_2pt}, four of which are displayed in Figure \ref{fig:fexp2}. 
For the subsequent analysis, we took the single-exponential results for the extracted ground 
state nucleon mass. Whilst we see a small discrepancy for this value obtained from double-exponential
fits, we find that, overall, the double-exponential fits largely confirm the single-exponential 
fits to be in the ground-state region. The observed discrepancy within the double-exponential fits (dependent on the fitting interval) is, in part, due to the contamination by higher excited states, which are 
not accounted for by the double-exponential fitting ansatz Eq.~\eqref{eq:ans_nn}.

For reference, we also show our previous results from \cite{Capitani:2015sba} in 
Table \ref{tab:fit_2pt} and note 
that for three ensembles (labelled E5, N6 and G8) we observe a discrepancy between the new 
single-exponential results and our previous results. This is due to the increased statistics 
on these ensembles, which allows us to better identify the contamination from excited states 
and hence provide a revised determination of the nucleon mass at later fitting intervals.

\section{AMA study of excited-state contamination in nucleon three-point functions}
\label{sec:3pt}
The ratios $R_A$, $R_S$ and $R_T$,
\begin{equation}
  R_{\mathcal{O}}(t,t_s) = g_{\mathcal O} + c_{\mathcal O} \big(e^{-\Delta t_s} + e^{-\Delta (t_s-t)}\big)
  + O(e^{-\Delta'(t_s-t)}), 
  \label{eq:r_o}
\end{equation}
with the target observable $g_{\mathcal O}$, mass difference $\Delta$ and unknown coefficient $c_{\mathcal O}$ 
suffer excited state contamination at finite $t$ and $t_s$. 
For the subsequent analysis, we used both the plateau method that assumes ground state dominance around the middle of each $t_s$ dataset and the summation method \cite{Gusken:1989ad,Capitani:2012gj,Jager:2013kha}.

\subsection{Axial charge}
\subsubsection{Plateau method}

\begin{figure}
\begin{center}
\includegraphics[width=0.48\linewidth]{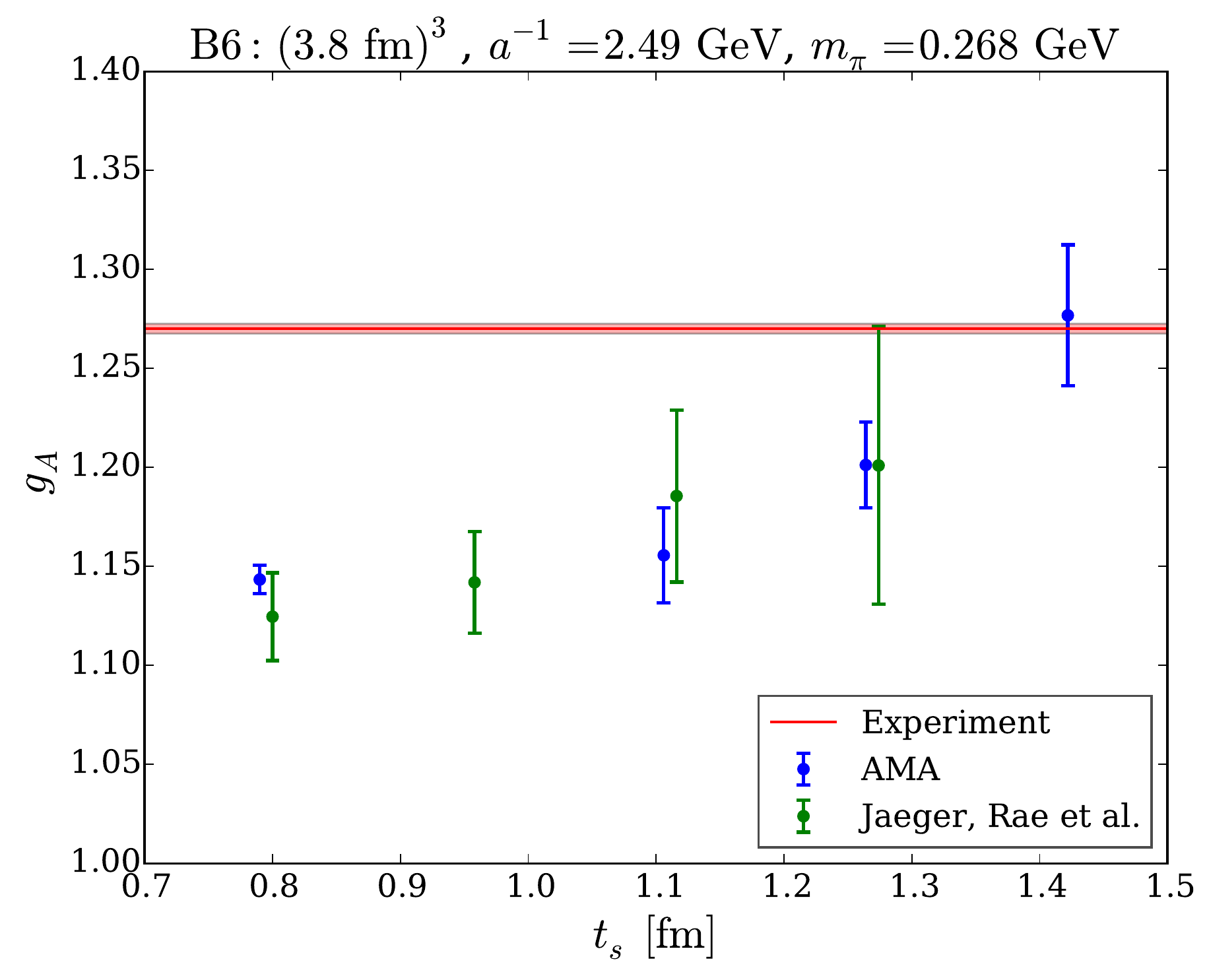}
\includegraphics[width=0.48\linewidth]{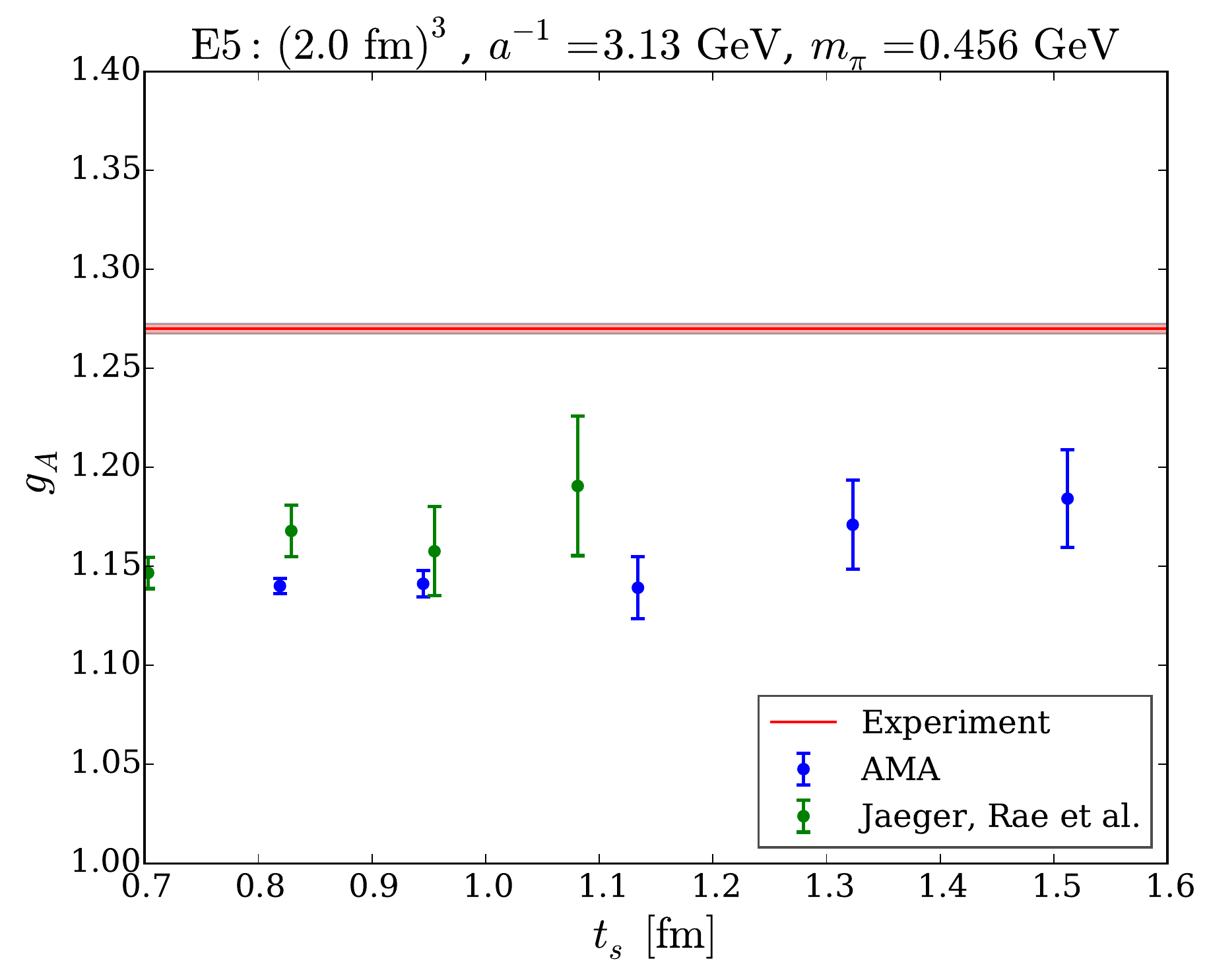}\\
\includegraphics[width=0.48\linewidth]{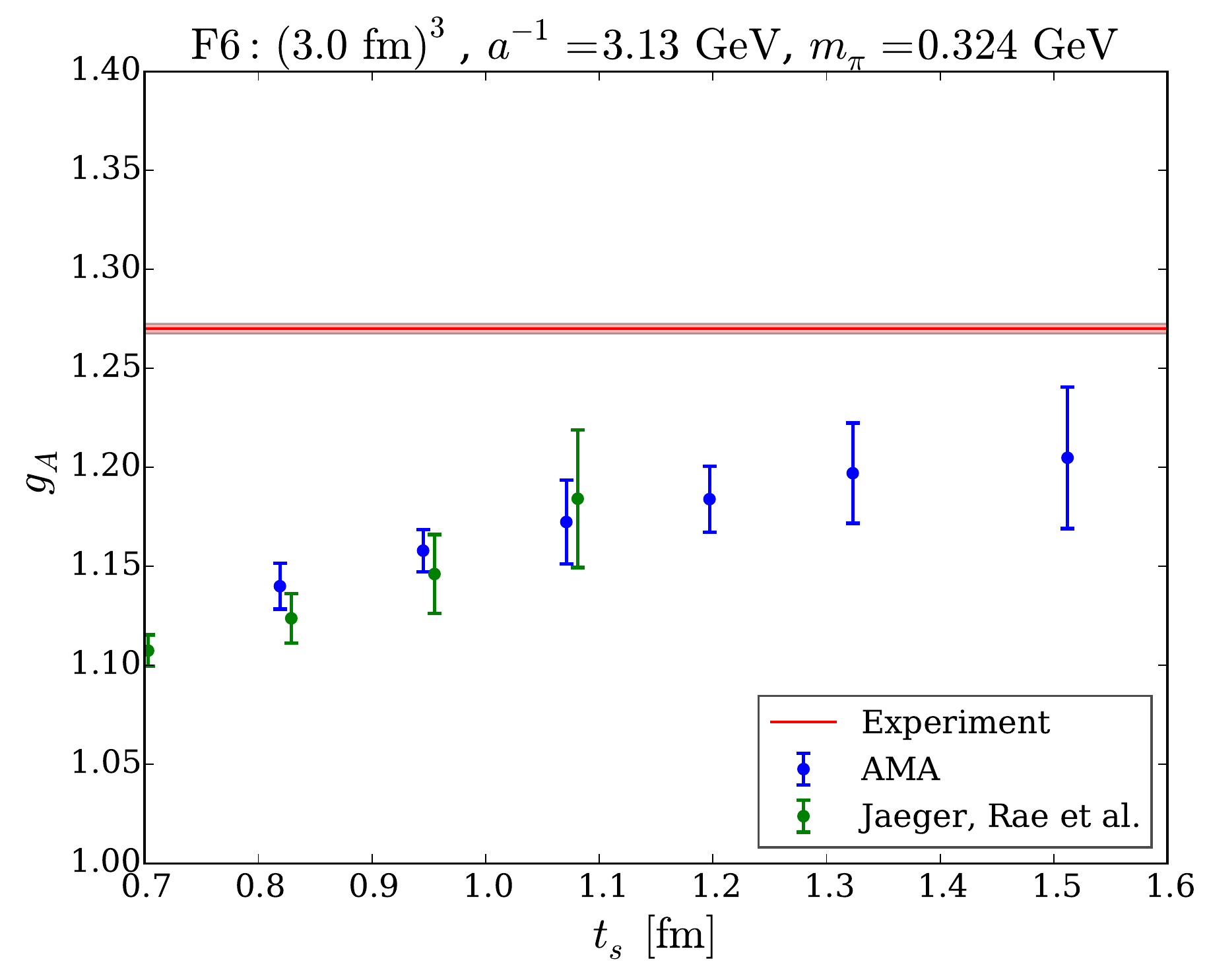}
\includegraphics[width=0.48\linewidth]{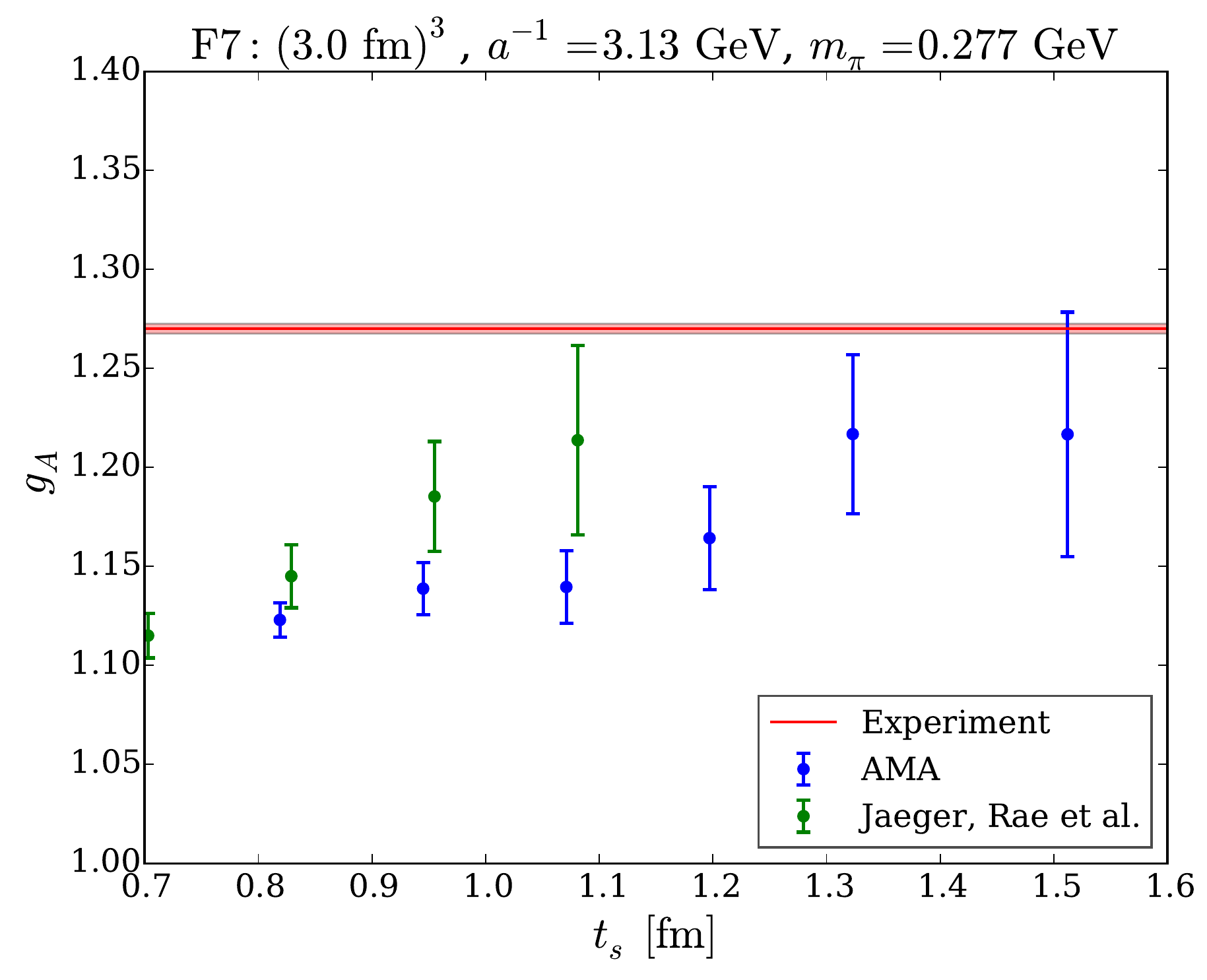}\\
\includegraphics[width=0.48\linewidth]{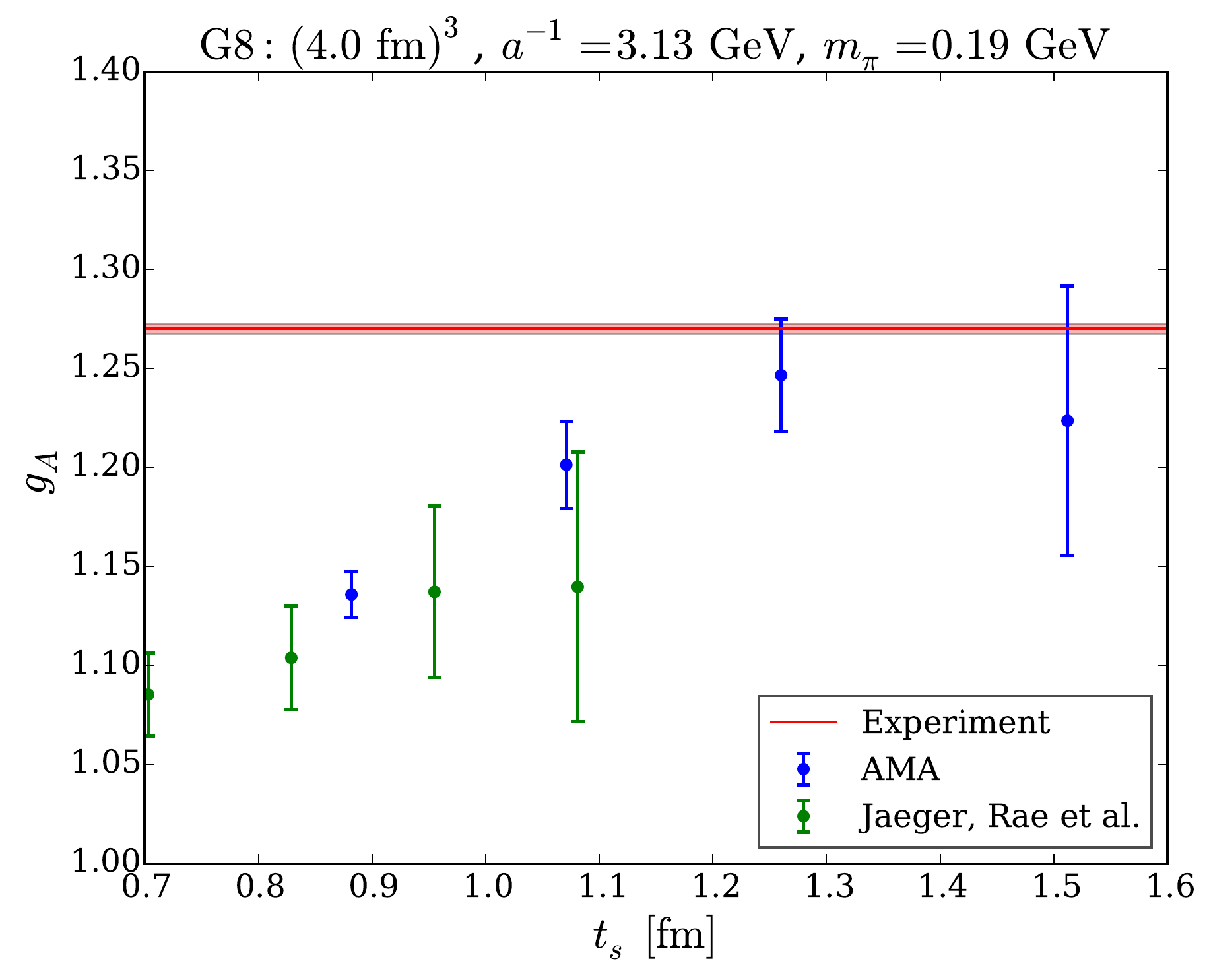}
\includegraphics[width=0.48\linewidth]{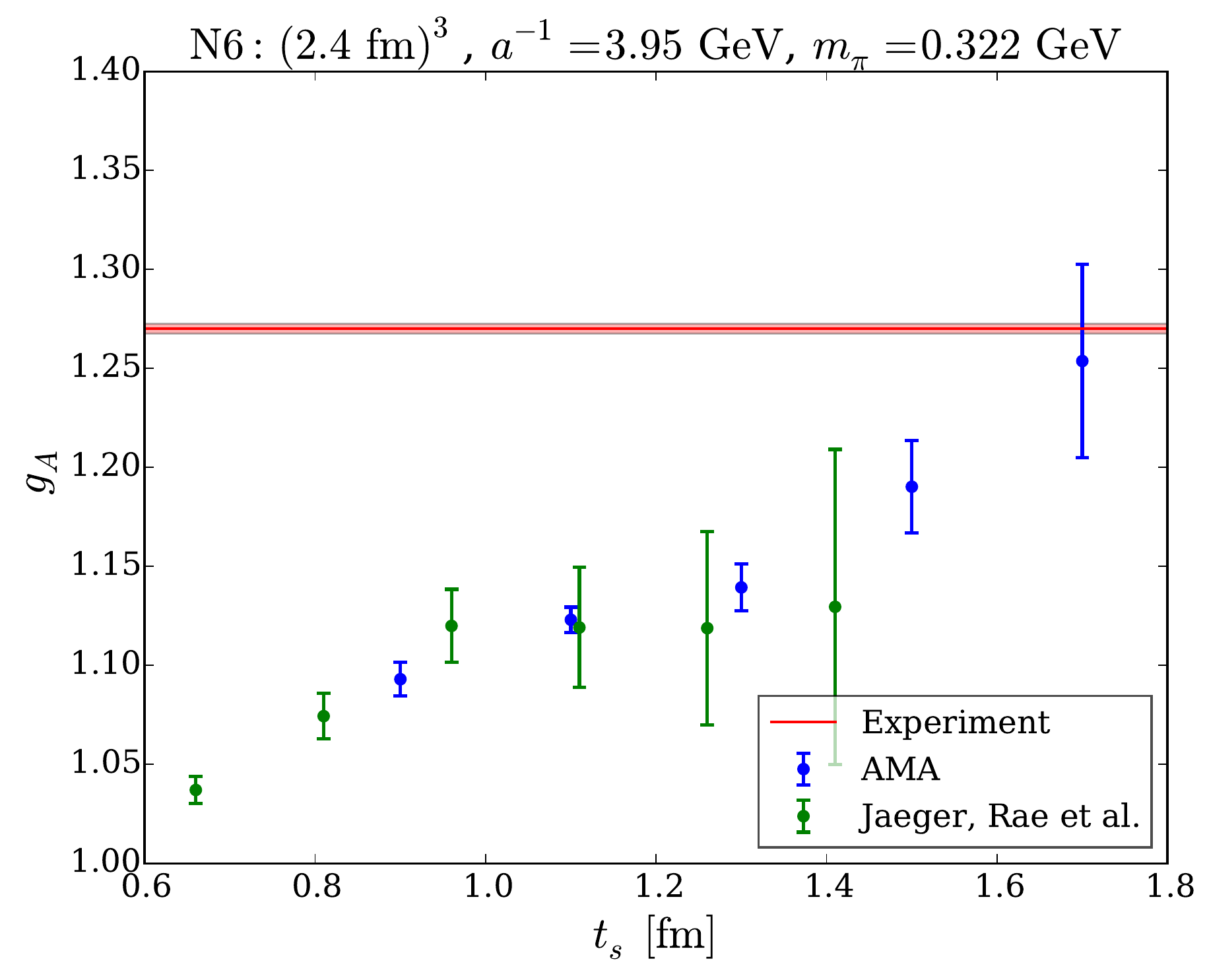}\\
\caption{$g_A$ as a function of $t_s$ from plateau fits. 
Green symbols denote our previous results 
\cite{Jager:2013kha} on the same configurations but without AMA.
These have been rescaled to account for the 
updated renormalisation factors in Table \ref{tab:zfact}.
}
\label{fig:ga_one_state}
\end{center}
\end{figure}

In Figure \ref{fig:ga_one_state} we show $g_A$ obtained
from plateau fits to the middle 4 points (3 for N6) of the ratio $R_A(t,t_s)$,
for each $t_s$. The ratio $R_A(t,t_s)$ typically resembles the example shown 
in Figure~\ref{fig:ga_sum} (right panel) for the F7 ensemble. 
This demonstrates a clear tendency for the plateau results to approach 
the experimental value from below as $t_s$ is increased and indicates a significant contribution from 
the excited states in Eq.~(\ref{eq:r_o}), which is especially pronounced for N6 (small lattice spacing) 
and G8 (small pion mass). The excited state contamination is significant at $t_s=1$ fm, which supports 
our previous findings \cite{Jager:2013kha} that the plateau method still suffers from substantial excited state contamination  
at $t_s=1.3$ fm and that 1.5~fm or more may be required. The plateau fit results at each 
$t_s$ for ensemble F7 are given in Table \ref{tab:ga_error_F7ts}, for which the extracted $g_A$ results can be clearly seen to 
approach the experimental value as the source-sink separation is increased. The results for the largest 
$t_s$ on each ensemble are summarised in Table \ref{tab:ga_error_ens}.

\begin{table}
\begin{center}
\caption{$g_A$ on the F7 gauge ensemble, $m_\pi=277$ MeV. Using the plateau method for each source-sink separation and the summation method result.}
\label{tab:ga_error_F7ts}
\begin{tabular}{ccc}
\hline\hline
$t_s/a$ & $t_s$~[fm] &  $g_A$ \\
\hline\hline
13 & 0.82  & 1.123(09)\\ 
15 & 0.95  & 1.139(13)\\ 
17 & 1.07  & 1.140(18)\\ 
19 & 1.20  & 1.164(26)\\ 
21 & 1.32 & 1.217(40)\\ 
24 & 1.51  & 1.217(61)\\ 
\hline
Sum (all $t_s$) & &1.218(48)\\ 
Sum ($t_s>0.9$~fm) & & 1.244(82)\\ 
\hline
Exp. & &  1.272(02)\\
\hline\hline
\end{tabular}
\end{center}
\end{table}

\begin{table}
\begin{center}
\caption{$g_A$ determined using the plateau method for
the largest source-sink separation, $t_s^{\mathrm{max}}$, for each gauge ensemble. The approximate pion and nucleon masses are given for reference.}
\label{tab:ga_error_ens}
\begin{tabular}{cccc}
\hline\hline
Label & $m_\pi$ [MeV] & $m_N$ [MeV] &$g_A$ plateau $t_s^{\mathrm{max}}$ \\
\hline\hline
A5 & 316 & 1160 & 1.255(35)\\
B6 & 268 & 1120 & 1.277(36)\\
\hline
E5 & 456 & 1350 & 1.184(25)\\ 
F6 & 324 & 1190 & 1.205(36)\\
F7 & 277 & 1150 & 1.217(61)\\ 
G8 & 190 & 1090 & 1.224(68)\\
\hline
N6 & 332 & 1130 & 1.254(49)\\ 
\hline
Exp. & 139 & 939 & 1.272(02)\\
\hline\hline
\end{tabular}
\end{center}
\end{table}

\subsubsection{Summation method}
Performing the summation $S_{\mathcal O}(t_s)$ parametrically reduces the
excited state contamination,
\begin{equation}
  S_{\mathcal O}(t_s) \equiv \sum_{t=1}^{t_s-1} R_{\mathcal O}(t,t_s)
  = d_1 + \big(g_{\mathcal O} + O(e^{-\Delta t_s})\big)t_s,
\end{equation}
and through determining $S_{\mathcal O}(t_s)$ for a number of $t_s$, 
the target observable may be obtained from the slope of a linear fit.

\begin{figure}
\begin{center}
\includegraphics[width=0.48\linewidth]{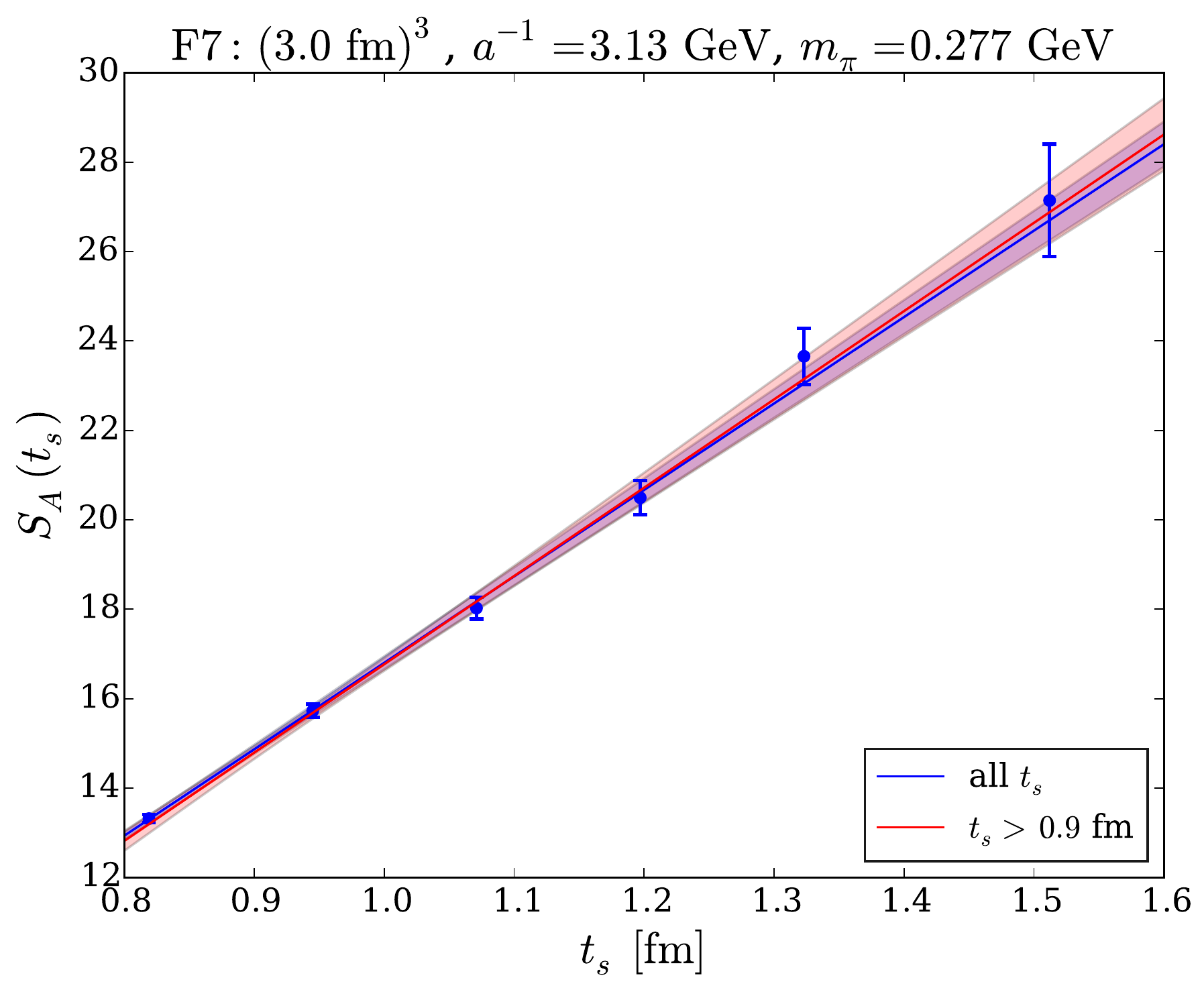}
\includegraphics[width=0.48\linewidth]{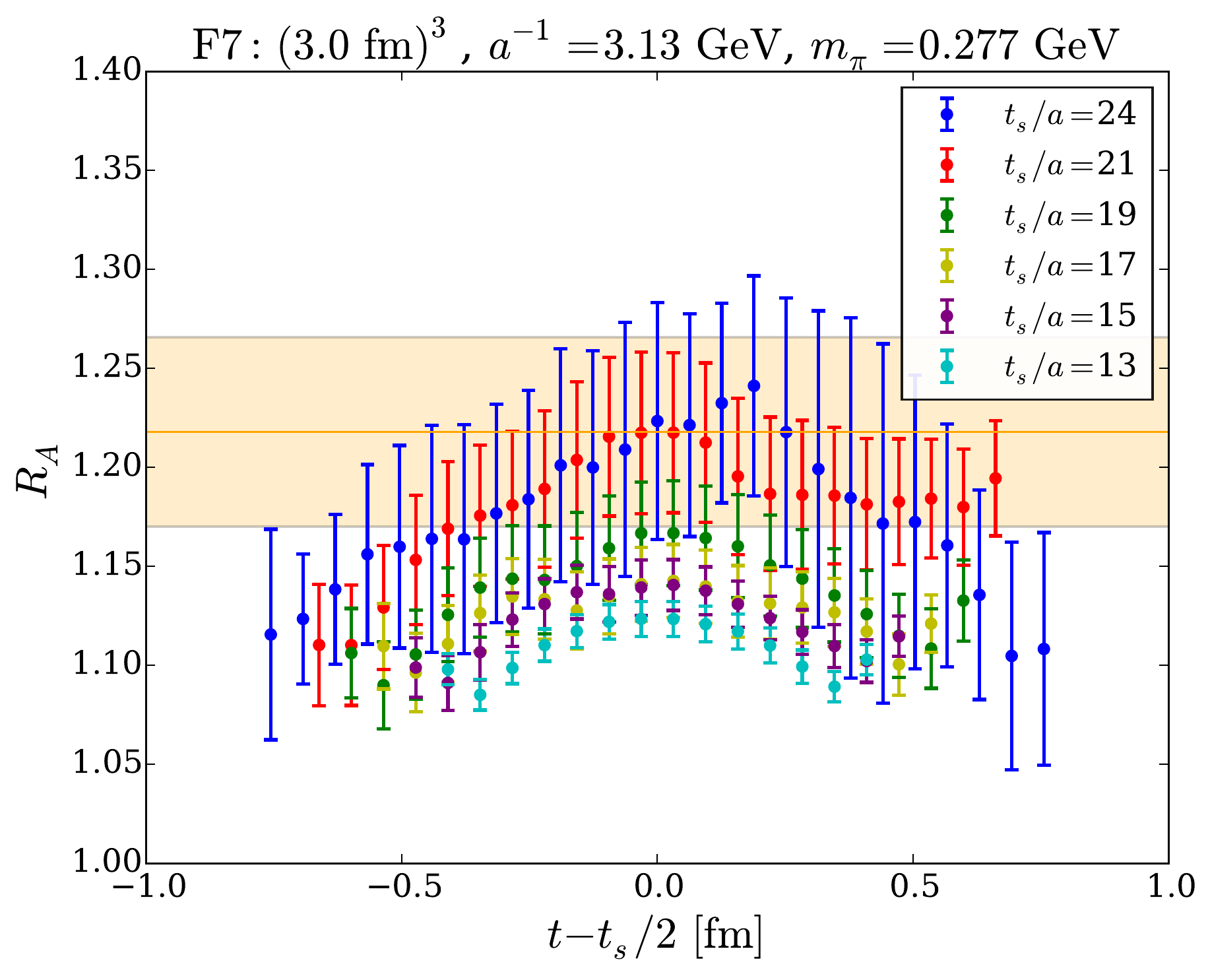}
\caption{Left panel: $S_A$ as a function of $t_s$ on the F7 gauge ensemble. 
The solid lines are linear fits to all $t_s$ points (blue) and only for points where $t_s>0.9$~fm (red).
The bands indicate the statistical error of the respective fit. 
Right panel: $R_A(t,t_s)$ for the nucleon axial charge as a function of $t-t_s/2$ with the summation result
for all $t_s$ overlaid.}
\label{fig:ga_sum}
\end{center}
\end{figure}

The $t_s$ dependence of the summed data points, including the
linear fits, for the F7 ensemble are shown in Figure \ref{fig:ga_sum} and 
the results are summarised in Table \ref{tab:ga_error_F7ts}. 
To check the dependence of the fit on the smaller $t_s$ points, 
the linear fits were performed for two intervals. 
One used all the $t_s$ points and the other used only the points where $t_s>0.9$~fm.
We observe no statistically significant discrepancy between the two fits and therefore we
quote the statistically more accurate result that incorporates all $t_s$ points.
 
\subsection{Scalar and tensor charge}

\begin{figure}[t]
\begin{center}
\includegraphics[width=0.48\linewidth]{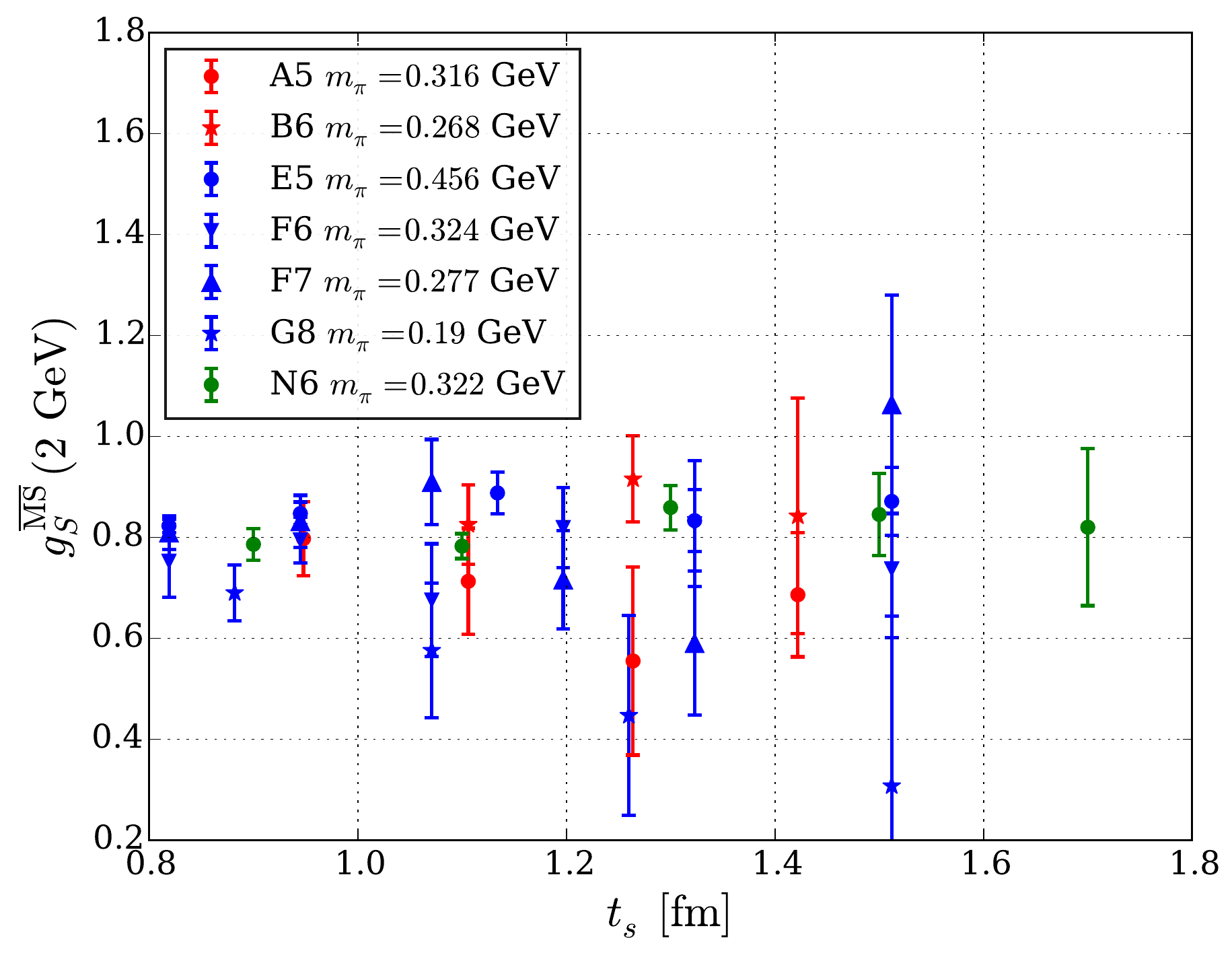}
\includegraphics[width=0.48\linewidth]{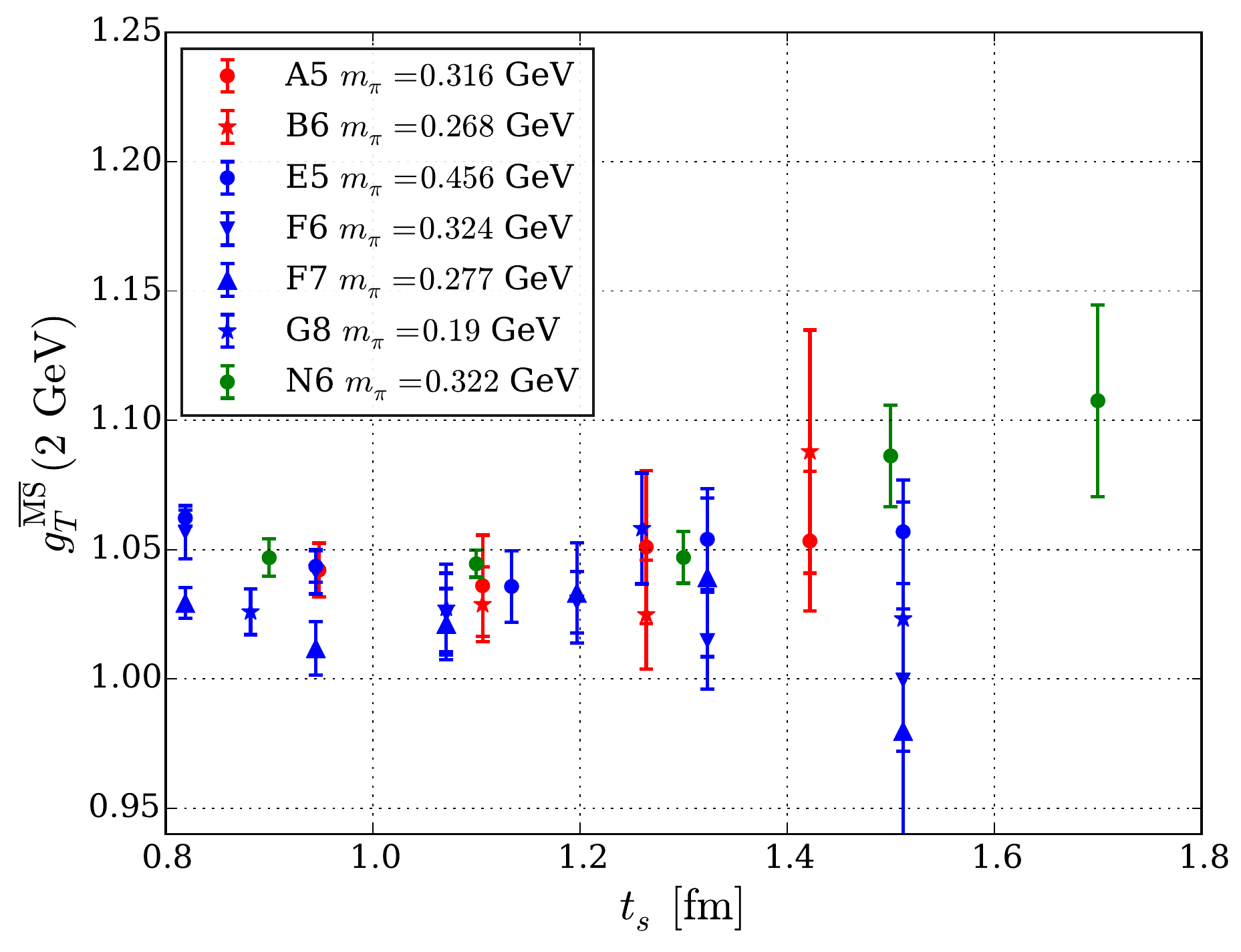}
\caption{Renormalised scalar charge $g_S$ (left panel) and $g_T$ (right panel) as a function of $t_s$ obtained from plateau fits.}
\label{fig:gs}
\end{center}
\end{figure}

\begin{figure}[t]
\begin{center}
\includegraphics[width=0.48\linewidth]{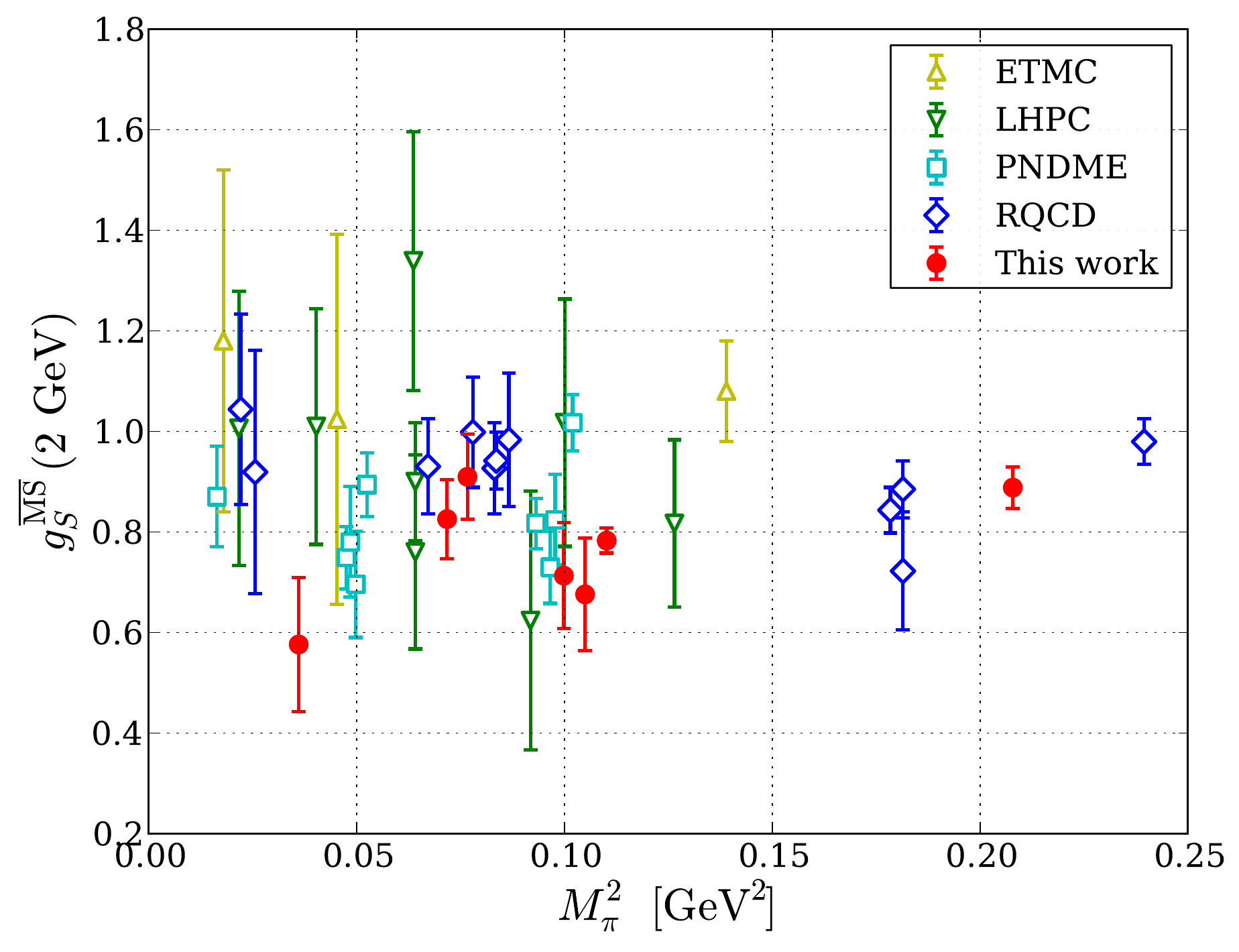}
\includegraphics[width=0.48\linewidth]{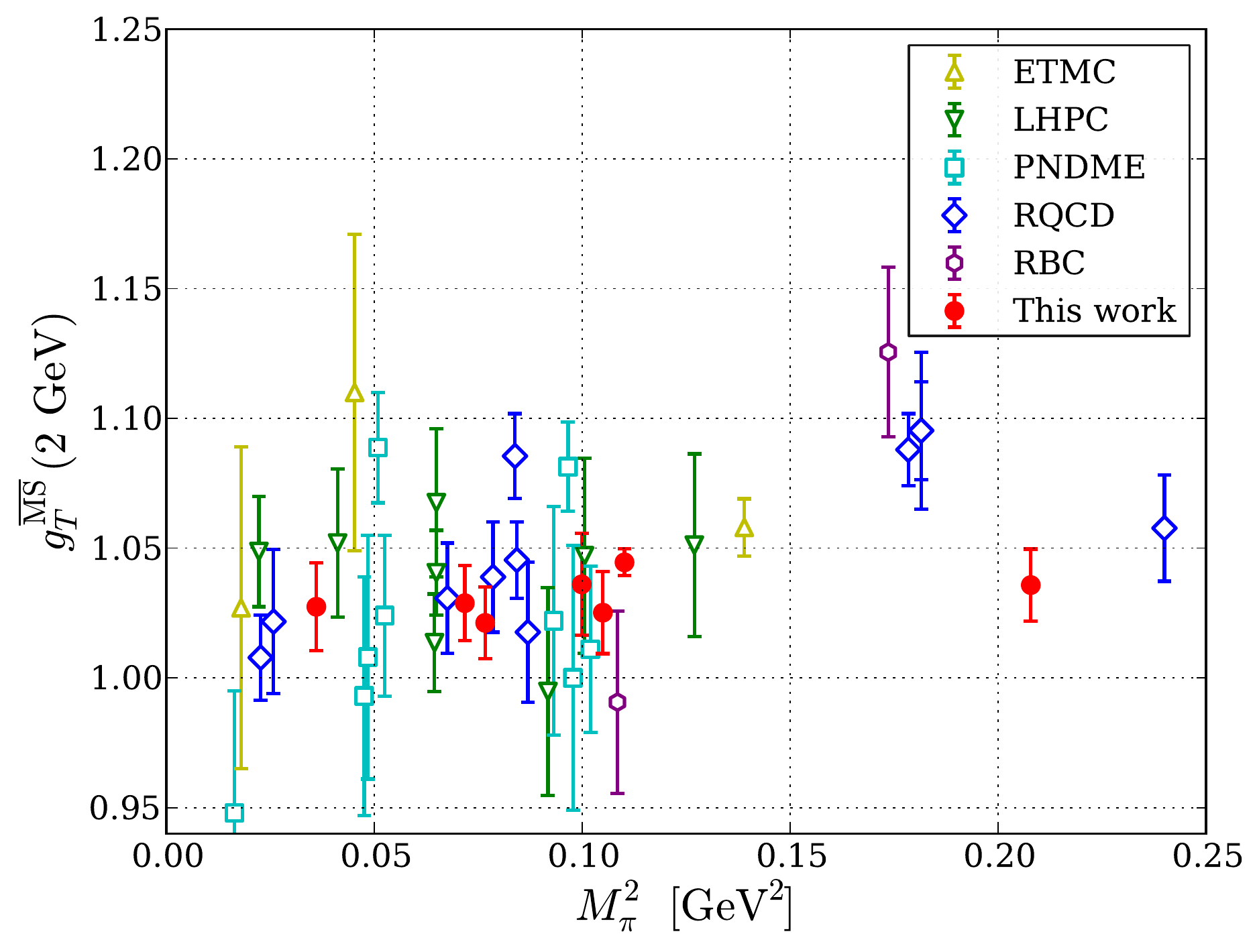}\\
\caption{Left panel: results for $g_S$ from ETMC \cite{Alexandrou:2014wca,Abdel-Rehim:2015owa}, LHPC \cite{Green:2012ej}, PNDME \cite{Bhattacharya:2013ehc,Bhattacharya:2016zcn}, and RQCD \cite{Bali:2014nma} (open symbols) compared to this work (filled symbols) as a function of $m_\pi^2$; right panel: results for $g_T$ from ETMC \cite{Alexandrou:2013wka,Abdel-Rehim:2015owa}, LHPC \cite{Green:2012ej}, PNDME \cite{Bhattacharya:2013ehc,Bhattacharya:2016zcn}, RQCD \cite{Bali:2014nma}, and RBC/UKQCD \cite{Aoki:2010xg} (open symbols) compared to this work (filled symbols) as a function of $m_\pi^2$. All results are quoted using a renormalisation scale of 2 GeV.}
\label{fig:cmp}
\end{center}
\end{figure}

\begin{figure}[t]
\begin{center}
\includegraphics[width=0.48\linewidth]{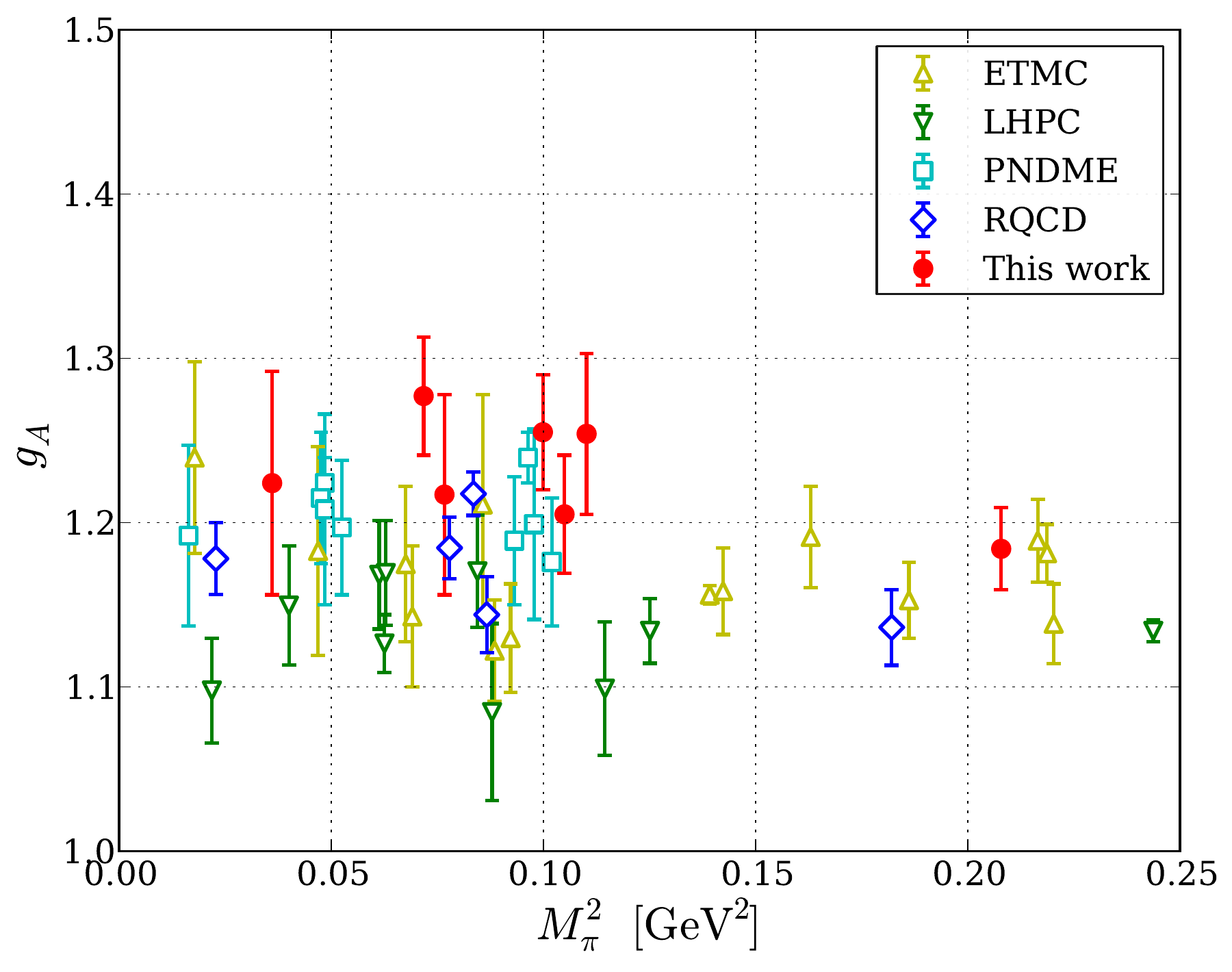}\\
\caption{Results for $g_A$ from ETMC \cite{Alexandrou:2010hf,Alexandrou:2013joa,Abdel-Rehim:2015owa}, LHPC \cite{Bratt:2010jn,Green:2012ud}, PNDME \cite{Bhattacharya:2013ehc,Bhattacharya:2016zcn}, and RQCD \cite{Bali:2014nma} (open symbols) compared to this work (filled symbols, plateau method only) as a function of $m_\pi^2$.}
\label{fig:cmp_gA}
\end{center}
\end{figure}

Figure \ref{fig:gs} shows the renormalised scalar charge $g_S$ 
and tensor charge $g_T$, extracted from the ratio of the two- and three-point functions 
defined in Eq.~(\ref{eq:R_S}) and Eq.~(\ref{eq:R_T}) and obtained from plateau fits 
(to the same intervals as for $g_A$).
In contrast to $g_A$, the dependence on $t_s$ 
is very mild with no evidence of excited state contaminations, 
even for a fine lattice spacing (N6) or for a light pion mass (G8). 
This echoes the behaviour seen by other groups, such as 
\cite{Bhattacharya:2013ehc,Bali:2014nma}. In Figure \ref{fig:cmp} we show a comparison
of lattice results for these quantities, where we use our plateau fit results for $t_s\sim1.1$ fm, which 
we believe to be reasonable due the absence of excited state behaviour in these quantities, 
as seen in Figure \ref{fig:gs}.
A similar comparison for the case of $g_A$ is shown in Figure \ref{fig:cmp_gA}, but we caution that the results obtained from the plateau method shown here should not be taken at face value due to the strong excited state contamination observed in the case of $g_A$, even when $t_s$ is as large as 1.3~fm.

\section{Summary and discussion}\label{sec:summary}

We have investigated the performance of all-mode-averaging (AMA)
\cite{Blum:2012uh,Blum:2012my,Shintani:2014vja}
when used in conjunction with the locally deflated SAP-preconditioned GCR
solver \cite{Luscher:2003qa,Luscher:2007se}
employed in the DD-HMC \cite{Luscher:2005rx,Luscher:2007es},
MP-HMC \cite{Marinkovic:2010eg} and
openQCD \cite{Luscher:2012av,openQCD} packages.
While the block decomposition that forms
the basis of the SAP preconditioner breaks the translation invariance of
the Dirac operator and thus limits our choices of source position for the
sloppy solver in the AMA method \cite{Shintani:2015aea}, we find that
AMA provides an increase in efficiency (as measured by computer time
expended to achieve a given relative statistical accuracy) by a factor
of around two compared to the standard method of averaging over multiple
source positions.

Using AMA with appropriately tuned parameters, we have investigated the
excited-state contamination of the axial, scalar and tensor charges of
the nucleon by measuring two- and three-point functions at large
source-sink separations $t_s\gtrsim 1.5$~fm with large statistics.
AMA enables us to achieve good statistical precision even at large $t_s$
with moderate computational effort. The results for the axial charge
contained in this paper represent an extension of our earlier
study \cite{Jager:2013kha} with statistics increased by a factor of 5--20.

From a comparison of different analysis methods (such as plateau fits
and the summation method), we are able to conclude that the ratio
from which $g_A$ is extracted still suffers from significant
excited-state contamination even at source-sink separations of around
$t_s\simeq 1.3$~fm, while the ratios for $g_S$ and $g_T$ are only weakly
affected.

In a recent study by the Regensburg group \cite{Bali:2014nma} using
similar configurations with $N_{\rm f}=2$ O($a$)-improved Wilson fermions,
a value for $g_A$ that was around 10\% below the experimental one was
obtained even at $m_\pi\simeq 150$~MeV, using source-sink separations
of $t_s\simeq 1$~fm.
Our estimate of $g_A$ obtained from $t_s\simeq 1.5$~fm in the mass range
$m_\pi\simeq 200-300$~MeV is consistent with experiment and shows now
significant pion mass dependence; we also do not find any evidence for
a sizeable finite-size correction \cite{Horsley:2013ayv} on our lattices
satisfying $m_\pi L\ge 4$.
It appears likely, therefore, that for $g_A$ source-sink separations
$t_s$ larger than 1.5~fm are required in order to avoid any
systematic uncertainty from excited state contamination.
This must be controlled before proceeding to an estimate of other
systematic uncertainties, for instance finite-size, pion-mass, and cut-off
effects.
Since we did not apply AMA in order to increase statistics to the entire
range of lattice spacings and pion masses available, we refrain from
performing a joint chiral and continuum extrapolation of $g_A$, $g_S$,
and $g_T$, to quote a result at the physical point.

The present study has demonstrated the feasibility of obtaining statistically
accurate results for nucleon structure observables from lattice QCD using
large source-sink separations with AMA, and we intend to perform further
studies of nucleon structure (including in particular the vector form
factors, where the suppression of excited-state effects appears to be of
crucial importance \cite{Capitani:2015sba} in order to address the
proton radius puzzle from first principles) using this method.
Related efforts on the $N_{\rm f}=2+1$ CLS configurations
\cite{Djukanovic:2015hnh} are under way.

\section*{Acknowledgments}
The authors acknowledge useful conversations with Benjamin Jäger.

Some of our calculations were performed on the
``Wilson'' and ``Clover'' HPC Clusters
at the Institute of Nuclear Physics, University of Mainz,
and the Helmholtz Institute Mainz.
We thank Dalibor Djukanovic
and Christian Seiwerth
for technical support.
The authors gratefully acknowledge the computing time granted
on the supercomputer Mogon at Johannes Gutenberg University
Mainz (www.hpc.uni-mainz.de).

This work was granted access to the HPC resources of the
Gauss Center for Supercomputing (GCS) at
Forschungzentrum Jülich, Germany,
made available within the
Distributed European Computing Initiative
by the PRACE-2IP,
receiving funding from the
European Community's Seventh Framework Programme (FP7/2007-2013)
under grant agreement RI-283493 (project PRA039).
The authors gratefully acknowledge the
Gauss Centre for Supercomputing (GCS)
for providing computing time to project HMZ21
through the John von Neumann Institute for Computing (NIC)
on the GCS share of the supercomputer JUQUEEN \cite{JUQUEEN}
at Jülich Supercomputing Centre (JSC).
GCS is the alliance of the three national supercomputing centres
HLRS (Universität Stuttgart),
JSC (Forschungszentrum Jülich),
and LRZ (Bayerische Akademie der Wissenschaften),
funded by the German Federal Ministry of Education and Research (BMBF)
and the German State Ministries for Research of Baden-Württemberg (MWK),
Bayern (StMWFK)
and Nordrhein-Westfalen (MIWF).

TR is grateful for the support through the DFG program SFB/TRR 55. 
ES is grateful to RIKEN Advanced Center for Computing and Communication
(ACCC).

We thank our colleagues within the CLS initiative for sharing ensembles.

\bibliographystyle{JHEP}
\bibliography{ref.bib}
\end{document}